%% file: main.tex
\documentclass[preprint,3p]{elsarticle}

\usepackage{lineno,hyperref}
\date{2018}
\journal{Pervasive and Mobile Computing}

\usepackage{amsmath}
\usepackage{amssymb}
\usepackage{amsfonts}
\usepackage{amsthm}
\usepackage{mathtools}

\usepackage{algorithm}
\usepackage{algorithmic}
\usepackage{array}

\usepackage[caption=false,font=footnotesize]{subfig}

\usepackage{graphicx}
\graphicspath{{./img/}}
 \DeclareGraphicsExtensions{.pdf,.jpeg,.png}

\usepackage[final]{changes}

\begin{document}

\begin{frontmatter}

\title{Energy efficient distributed analytics at the edge  of the network for IoT environments}

\author[iit]{Lorenzo Valerio\corref{cor}}
\cortext[cor]{Corresponding authors}
\ead{l.valerio@iit.cnr.it}
\author[iit]{Marco Conti}
\ead{m.conti@iit.cnr.it}
\author[iit]{Andrea Passarella}
\ead{a.passarella@iit.cnr.it}

\address[iit]{Insitute of Informatics and Telematics, National Research Council, Via Moruzzi 1, Pisa. Italy}

\begin{abstract}
Due to the pervasive diffusion of personal mobile and IoT devices, many ``smart environments'' (e.g., smart cities and smart factories) will be, generators of huge amounts of data. 
Currently, analysis of this data is typically achieved through centralised cloud-based services.  However, according to many studies, this approach may present significant issues from the standpoint of data ownership, as well as wireless network capacity.   
In this paper, we exploit the fog computing paradigm to move computation close to where data is produced. We exploit a well-known distributed machine learning framework (Hypothesis Transfer Learning), and perform data analytics on mobile nodes passing by IoT devices, in addition to fog gateways at the edge of the network infrastructure. We analyse the performance of different configurations of the distributed learning framework, in terms of (i) accuracy obtained in the learning task and (ii) energy spent to send data between the involved nodes. Specifically, we consider reference wireless technologies for communication between the different types of nodes we consider, e.g. LTE, Nb-IoT, 802.15.4, 802.11, etc. Our results show that collecting data through the mobile nodes and executing the distributed analytics using short-range communication technologies, such as 802.15.4 and 802.11, allows to strongly reduce the energy consumption of the system up to $94\%$ with a loss in accuracy w.r.t. a centralised cloud solution up to $2\%$.

\end{abstract}


\end{frontmatter}

\input{intro}

\input{related}

\input{scenario}

\input{methodology}

\input{results}
\input{complexity}
\input{conclusions}
\section*{Acknowledgment}

This work is partly funded by the European Commission under the H2020
REPLICATE (691735) , SoBigData (654024), and AUTOWARE
(723909) projects.

\section*{References}

\bibliography{bibliography}

\end{document}

%% file: intro.tex
\section{Introduction} \label{sec:intro}

There is a unanimous consensus in the research and industry communities about the fact that IoT applications will account for a large share of the mobile data generated in the Internet~\cite{Cisco:sp}. Many reference market analyses (e.g.~\cite{McKinsey}), show that the adoption of IoT technologies will result in a huge economic impact in many sectors, well beyond ICT alone.
As most of the value of IoT applications will come from the analysis of the data generated by IoT devices, the research area of IoT data analysis and management is a very exciting one.

The common trend in many current architectures~\cite{Borgia20141} is to transfer IoT data from the physical locations where they are generated to some global cloud platform, where knowledge is extracted from raw data and used to support IoT applications. This is the case, among others, of the ETSI M2M architecture~\cite{ETSI-arch}. However, there are concerns whether this approach will be sustainable in the long run. The projections of the growth of the number of deployed IoT devices are exponential over the next years~\cite{McKinsey}.  Mobile data traffic will grow at a compound annual growth rate (CAGR) of 47 percent from 2016 to 2021, reaching 49.0 Exabytes per month by 2021~\cite{Cisco:sp}. Together with data generated by personal users' devices, this is likely to make the amount of data generated at the edge of the network huge, making it impractical or simply impossible to transfer them to a remote cloud platform at reasonable costs.
In addition, data might also have privacy and confidentiality constraints, which might make it impossible to transfer them to third parties such as global cloud platform operators. 
For example, in the manufacturing domain, data analytics is one of the cornerstones of the Industry 4.0 or Industrial Internet \cite{Kagermann:2013aa} concepts. However, most of the time industries will not be willing to move their data to some external cloud provider infrastructure, due to confidentiality reasons. On the other hand, they might not have the competencies and resources to build and manage a private cloud platform.
Moreover, real-time delay constraints might require that data elaboration or storage is performed at the edge, i.e., close to where it is needed, rather than in remote data centres. Last but not least, as we show in this paper, sending data from IoT devices to remote data centres might result in very high energy consumption, even using wireless technologies designed for low power devices, such as NB-IoT.
These trends and needs push towards a decentralisation of cloud platforms towards the edge of
the network, where the  paradigms of Edge computing, such as Fog~\cite{Bonomi:2012aa}, Mobile Edge
Computing~\cite{Lopez:2015aa} and Cloudlets\cite{Satyanarayanan:2009aa}, can address these problems.


In this paper, we follow this approach and study the behaviour of a distributed learning solution based on Hypothesis Transfer Learning (HTL). 
In general, with HTL, instead of training a model on the whole training set in a centralised way, multiple parallel models are trained on disjoint subsets, and then the partial models are combined to obtain a single final model. We already have successfully applied HTL to the case of distributed learning in IoT environments in~\cite{Valerio:2016aa} and \cite{VALERIO201746}, where we have presented an activity recognition solution, where 
we have shown that such solution is able to drastically cut the network traffic required to perform the learning task, with an affordable reduction of accuracy with respect to a conventional solution where data is transferred on a global cloud platform.

Contrarily to~\cite{Valerio:2016aa,VALERIO201746}, in this paper we consider that data, generated by IoT devices, need to be moved either to an edge gateway (and possibly to the cloud) or to a number of mobile nodes passing by the IoT devices. Both the gateway and the mobile nodes take the role of Data Collectors, and HTL is used to compute and exchange partial models between them. 
The main focus of this paper is on evaluating the interplay between the configuration of the distributed learning scheme and the involved network costs. This primarily depends on the assumptions about the spatio-temporal distribution of the nodes, and on the transmission costs for the various considered technologies.
Specifically, we consider realistic wireless technologies for the communication between the nodes involved in the scenario, and we study the performance of different configurations of the HTL process in terms of the trade-off between the energy saving obtained by keeping data at the edge (and thus using short-range instead of long-range wireless technologies), and the loss of accuracy of the HTL scheme with respect to a conventional centralised learning scheme that can work on the entire set of data generated by all IoT devices.
We compare different HTL configurations between them, and with a centralised solution where all data are sent to a remote cloud platform for processing. 
We compare different strategies of data collection, i.e. we go from the centralised solution in which data is collected by the edge server through cellular communication to using exclusively mobile nodes, in which devices collect all the data and perform the learning.  
We consider the most relevant wireless communication technologies, i.e., NB-IoT, LTE, 802.15.4 and 802.11. Precisely, we assume that IoT devices communicate with an edge gateway through energy saving cellular technologies (NB-IoT) and with mobile nodes using 802.15.4. Moreover, Mobile devices can communicate with each other through 802.11 and with the edge gateway using LTE.

This paper extends~\cite{Valerio:2016ab}, as we take into account and evaluate in detail the impact of the specific wireless technologies used for communication, and we also consider the case where mobile nodes may lose data from IoT devices 
(while in ~\cite{Valerio:2016ab} data collectors are only static nodes). 

Changing the dataset would change the accuracy of our distributed learning scheme with respect to a centralised solution, which is something we have analysed extensively in [1,2,3] considering multiple datasets.
However, for the specific focus of this paper, we think that focusing on one dataset already would be enough, as, thanks to our prior results, we can expect qualitatively similar results when the same analysis is performed on other datasets.
We clarified this point in Section 1 of the paper. 

To test the considered alternatives in a realistic setting, we selected a dataset related to an environmental monitoring application.\footnote{For the specific focus of this paper, we think that focusing only on one dataset would be enough, as, thanks to our prior results in \cite{VALERIO201746,Valerio:2016aa,Valerio:2016ab}, we can expect qualitatively similar results when the same analysis is performed on other datasets.} Precisely, the learning problem we consider is to learn a classifier able to predict the type of trees covering a delimited area of forest. 
We simulate an iterative data collection process interleaved by sessions of distributed learning on the newly collected data.  However, for the specific focus of this paper, we think that focusing on one dataset already would be enough, as, thanks to our prior results, we can expect qualitatively similar results when the same analysis is performed on other datasets.

Our results show that it is possible to save up to $94\%$ of energy in the system (mostly related to the data collection part) without relying on the edge servers with up to $2\%$ of accuracy degradation with respect to a centralised cloud-based solution. Moreover, the computational burden of part of our solution can be significantly reduced with a negligible impact in terms of accuracy performance. 

The rest of the paper is organised as follows. Section~\ref{sec:related}
presents related work.
In Section~\ref{sec:statement} we introduce the problem and the reference scenario we consider in the paper. Section~\ref{sec:methodology} describes the HTL-based distributed learning solutions used in our experiments. 
Sections~\ref{sec:settings} 
describes the dataset and the performance metrics we used to perform the analysis presented in Section~\ref{sec:results}. In Section \ref{sec:complex} we discuss the computational complexity of our solution. Finally,
Section~\ref{sec:conclusions} concludes the paper.

%% file: related.tex
\section{Related work}
\label{sec:related}
The distributed learning problem can be approached in many different ways. One typical approach is represented by the so-called ``ensemble methods'' such as bagging, boosting and mixtures of experts~\cite{Breiman:aa, Freund:1997aa, Freund:1997ab, Jacobs:1991aa}. Typically, these methods randomly split a dataset into smaller portions and allocate them to different classifiers that are independently trained. The individual classifiers are subsequently aggregated, sometimes after an additional training phase or with other techniques based on feedbacks coming from the training phase. Although these approaches are very powerful, they assume that the training set is available to a single coordinating processor.

Another very promising set of powerful solutions able to process huge amounts of data is represented by deep learning techniques~\cite{Bordes:2005aa, Dean:2012aa, Coates:2013aa}. These techniques are nowadays widely used to solve many complex tasks.  Differently from our solution, these approaches do not target knowledge extraction where data have privacy constraints, or when network overhead should be minimised.

Other works propose fully distributed and decentralized learning algorithms. To the best of our knowledge, the more relevant with respect to our reference scenario are the following.  In~\cite{navia2006distributed} authors present a distributed version of Support Vector Machine (SVM).  Authors of~\cite{Georgopoulos20142} propose a distributed learning algorithm for arbitrarily connected machines.  Another similar solution presented in~\cite{Scardapane2015271} propose two distributed learning algorithms for random functional link networks. Both are iterative solutions that in order to converge to a model have to repeatedly exchange their partial models' parameters. 

Other approaches that are closer, at least in terms of methodology to the one proposed in this paper include \cite{Konecny:2016aa}, where the authors propose a framework called Federated Learning whose purpose is to accomplish a distributed learning task on several mobile devices without exporting data from them. Briefly, in their approach, they train a neural network by executing the learning algorithm on each device and collecting in a centralised fashion the updates of the model. It proves to be very effective but, differently from our paper, they do not take into account aspect connected to the amount of traffic generated by the learning task. 

Approaches based on distributed learning are emerging in the pervasive networking literature, e.g., to address the activity recognition problem.  For example, in \cite{Oquab:2014aa,Yosinski:2014aa} several transfer learning-based approaches exploiting deep learning solutions are applied to the activity recognition problem. However, their main focus is on finding the most effective way to transfer layers of a neural network in order to exploit that knowledge to ease the training in other similar domains. 

Finally, in \cite{Valerio:2017aa} we have proposed an analytical solution that exploits a reference gradient-based learning algorithm (i.e. SVRG) to identify the optimal trade-off between accuracy and network overhead. Precisely in that we find the optimal operating point of data aggregation on the nodes that guarantee a certain target model accuracy. Differently, in this paper, we consider an approach based on the Hypothesis Transfer Learning framework in which entire models are moved between nodes instead of single updates of a single model, computed in separate locations. 

The purpose of all these solutions is to learn a model whose accuracy performance is close, or at least comparable, with a centralized algorithm with access to the complete dataset. Although the accuracy is a crucial evaluation term for such kind of solutions, also the energy consumption triggered by those algorithms is fundamental, if we consider a network constrained scenarios like the IoT. In particular, in this paper, we are interested in evaluating the behaviour of the Hypothesis Transfer Learning framework in a context where realistic network related constraints are imposed.

%% file: scenario.tex
\section{Problem statement and system assumptions}
\label{sec:statement}
\begin{figure}[ht]
    \centering
    \subfloat[Data collection]{
    \includegraphics[width=.5\textwidth]{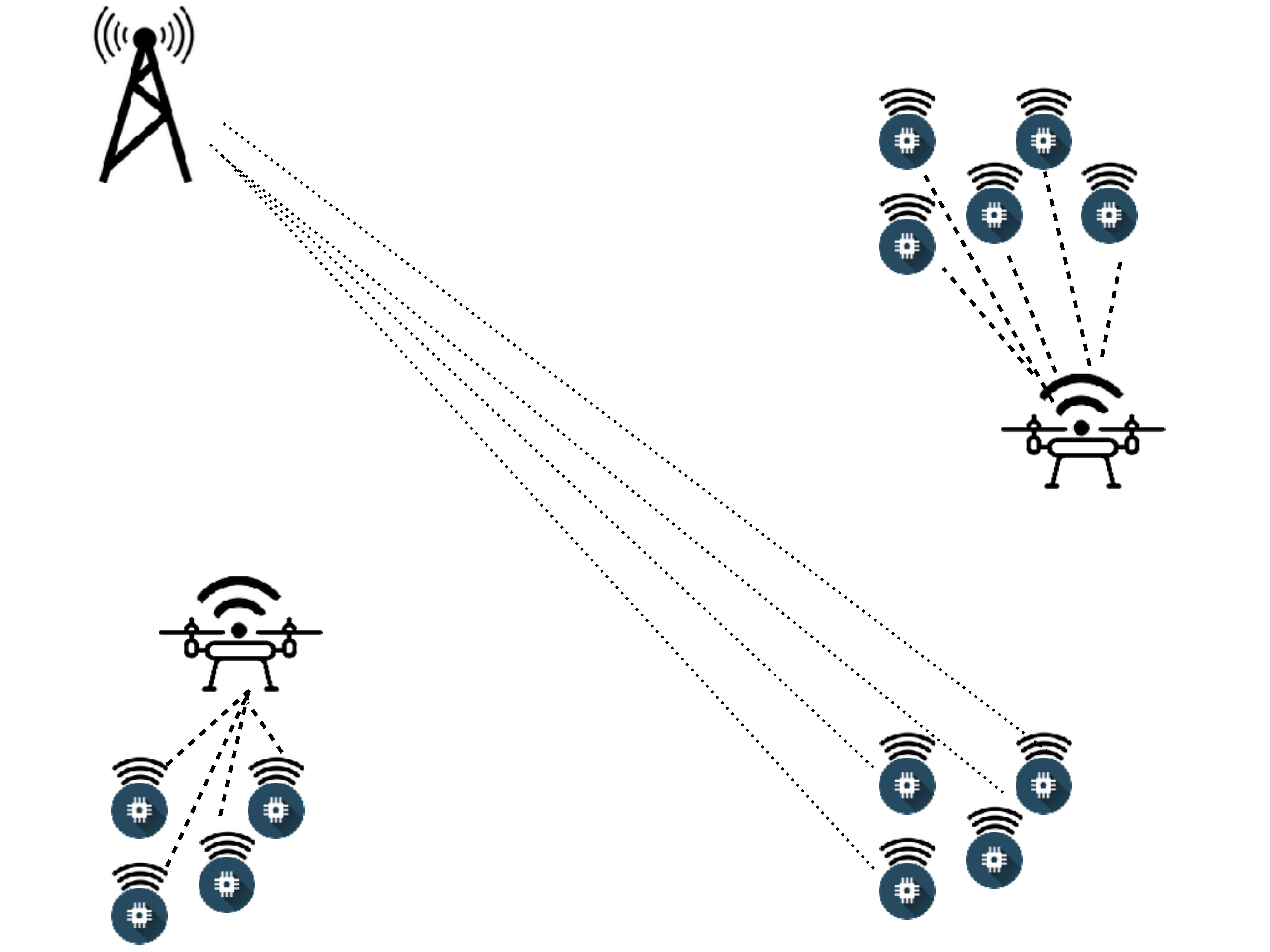}
    \label{fig:sc_data_coll}
    }
    \subfloat[Dist. data processing]{
    \includegraphics[width=.5\textwidth]{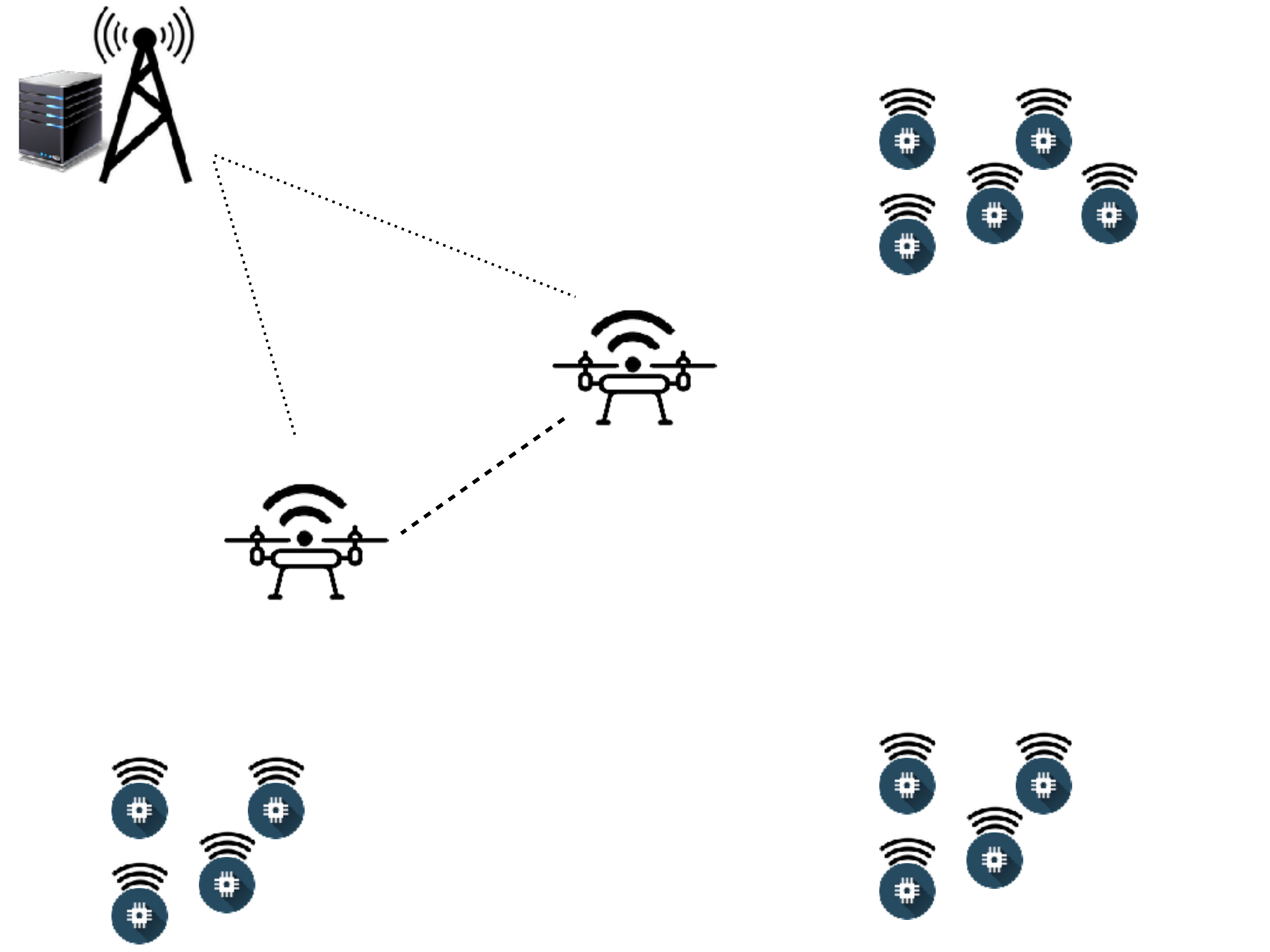}
    \label{fig:_sc_data_proc}
    }
    \caption{Reference scenario}
    \label{fig:scenario}
\end{figure}

In this section we describe the reference scenario we consider in this paper, as shown in Figure \ref{fig:scenario}.
We assume a wide area  that has been disseminated with several types of IoT devices, such as temperature sensors, humidity sensors, etc. 
These sensors are battery powered and they have limited computational and storage capacity. Since they are resource constrained, all the data they collect from the environment has to be transmitted to other more capable devices, in order to be processed. 
In this paper, we consider a slotted data collection process, that is, the sensors sense the environment and store the data for a fixed amount of time. 
At the end of a generic slot, they must transmit the data (within a certain deadline) in order to clean their memory and make room for the next session of data collection. 
In our scenario, the sensors have two ways for transmitting the collected data. In the first one, if a mobile device, that from now on we will call SmartMule (SM), is in the communication range of the sensors,  it collects data from them through a short-range wireless technology. Note that a SmartMule can be any kind of mobile device capable to collect, process and analyse data, e.g.  an SM can be a drone equipped with a chip-set appropriate for processing data, or a smart-phone, according to the specific sensing problem. The second one is used when no SM has come in the communication range of the device by the stated deadline. In that case a long range transmission occurs in order to send data to an edge server (ES)  located at the edge of the network. For simplicity, we will term both SMs and ES as Data Collectors (DC). From a technological point of view, for long range transmissions, sensors can use one of the energy saving cellular technologies, e.g. NB-IoT~\cite{nbiot:2016aa}, while for short/medium range transmissions they exploit a communication protocol based on the IEEE 802.15.4 standard, e.g. Zigbee~\cite{zigbee:aa}. Figure~\ref{fig:sc_data_coll} shows the data collection possibilities, through short range technologies (dashed lines) or long range ones (dotted lines).

At the end of each collection session, all DCs  collaboratively learn a model (through HTL) based on the newly acquired data. This model is used to update the model elaborated until the previous time slot. Therefore, over time DCs elaborate a model, which is incrementally refined through the data collected after each collection slot.\footnote{The main assumption here is that the time between consecutive batches should be significantly longer than the time required to train the model based on the data in each batch. If this is not the case, the entire approach considered in the paper would not be useful, as the model would be constantly "behind schedule" with respect to data generation rate. In such cases, either the application would downgrade its sampling rate, or mules should be improved in order to compute more quickly. However, this would be a dimensioning problem that is beyond the aim of this paper.} 
It is worth noting that in this scenario data has not a locality constraint, i.e. once arrived on the first SM, it can be further aggregated on other SMs or even on the ES, according to the data management policy associated to the learning process.
We assume that data can either be elaborated directly on the DC which collect them from the IoT devices, or can be exchanged and grouped at a smaller number of DCs, which are the sole responsible for computing the model. This allows us to have flexibility on the number of DCs, and on the locations where models are elaborated. We will study the impact of this configuration parameters in Section \ref{sec:results}.
In our scenario, we consider that devices can communicate with each others in two ways. One way is through  cellular technology such as LTE or LTE/A. In this case we are assuming that SMs are far from each others (long range communication). Conversely if the devices are close enough, they can communicate through a  short/medium range technology, such as WiFi or Bluetooth. \footnote{The concept of ``close enough'' may vary depending on the specific technology considered.} 

In order to make our scenario more realistic we consider that the number of SMs roaming in the area for collecting data may vary from slot to slot. In addition, not only the number of SMs but also the amount of data they collect might be different from each other. 
To this end, the number of SMs present in the area during a single collection session is drawn from a Poisson distribution. 
In addition, we consider that SMs have a different mobility patterns in the sense that some SM may have the chance to visit a greater number of sensors than others, leading to an uneven amount of data collected on DCs. Specifically, we model the number of data collected by DCs during a session with a Zipf distribution. In order to allocate data to DCs we use the following procedure. To each  DC in the area we assign an ID which also corresponds to a raking value from 1 to N, where 1 is the highest ranking and N is the lowest one (note that N is the total number of DCs in the area, drawn from the Poisson distribution). According to the Zipf's law each ranking value has a different probability of being drawn, therefore, in order to assign data to DCs, for each datum we randomly draw one of the N DCs' IDs. The result of such process is that the distribution of data on DCs is unbalanced, i.e. the number of data on each device and their internal distribution is not uniform. Moreover, we assume that the data collected by DCs are homogeneous, i.e., all the sensors positioned across several separate physical locations,  collect data (possibly in separate time instants) belonging to the same generating phenomenon. Therefore, although mules in separate collection windows might have only a partial view of the all available data, in the long run they have access, on average, to all the available information, either in form of raw data or in form of aggregate partial models.

In this paper the learning task that all the DCs  have to solve in a collaborative and distributed way is a classification problem. Specifically, we assume that each data partition located on a DC is a training set used to train a local classifier. The features of the local classifiers are exchanged with all the other DCs and used by each of the them to refine their own local classifier. These refined models are finally aggregated all together in order to combine their knowledge into a single ``global'' model, which is the one updated after each session. We point out that the model we learn from data is a linear classifier, that is the one of the simplest and the most lightweight models that can be learnt in terms of number of parameters. Nevertheless, its simplicity does not affect the generality of our analysis, because selecting a different model would affect only the quantity of data exchanged between  DCs and not the ranking of the several configurations that we present in the rest of the paper. 

Our purpose is to evaluate the performance of such system both in term of accuracy of the model learnt  and the associated cost, defined as the energy spent for the wireless communications~\footnote{In this paper we do not include in our analysis the energy count used for the computation. We consider it as an orthogonal problem that deserves a dedicated study because it depends on many factors such as, among others, the specifics of the reference computational architecture.}.
Precisely, our aim is to evaluate the impact that the considered communication technologies have on the cost of each part such system and to identify, depending on the specific operational conditions of the system,  what are the most energy efficient configurations of the HTL algorithm in terms of where data has to be moved across DCs, for a certain target accuracy of the model.

%% file: methodology.tex
\section{Distributed Learning through Hypothesis Transfer Learning}
\label{sec:methodology}
In this section we briefly  describe the distributed learning solution we use in this paper, based on the Hypothesis Learning Framework (HTL).

HTL methods are machine learning algorithms through which it is possible to exploit the knowledge acquired by a model $m_1$ trained on a certain set of data $D_1$ to ease the training phase of a model $m_2$, trained on a set of data $D_2$, different from $D_1$. The purpose is to improve the accuracy of $m_2$ using the knowledge contained in $m_1$.

 This kind of solution is typically used in situations in which  the size of $D_2$ is much smaller than the size of $D_1$. In fact in this situations, without a technique such as HTL, the accuracy of $m_2$ would be strongly affected by the small amount of data available to train the model. 

In this paper, in order to perform the analysis described in Section~\ref{sec:statement}, we exploit the distributed learning solution proposed in \cite{Valerio:2016aa,VALERIO201746} whose core learning mechanism is based on GreedyTL, i.e. one of the reference solutions in the HTL literature~\cite{OrabonaGreedyTL}. Note that we do not provide all the mathematical details of GreedyTL. The interested reader should refer to the original paper~\cite{OrabonaGreedyTL}. 
Instead, we describe the key points of two distributed learning procedures. The first one is the distributed learning solution proposed in \cite{Valerio:2016aa,VALERIO201746}, that here we adapt to the scenario presented in Section~\ref{sec:statement}. The second one is a modified version of the former, where the key difference is that the second scheme significantly reduces the amount of data exchanged between the DCs, as we explain in the following of the section.

From now on the first scheme is called \emph{All-to-all HTL } (A2AHTL), while the second one \emph{StarHTL} (SHTL).
The basic assumption, that is valid for both A2AHTL and SHTL, is that each of the $L$ DCs holds a local set of data $D_i,\mathrm{with}\ i=1,\dots,L$ and each $D_i$ is different from the others, i.e. $D_i\cap D_j=\emptyset, \forall i,j=1,\dots,L$ with $i\neq j$. Note that, the local datasets differ from each others not only for the content but also in terms of size, i.e. $|D_i| \neq |D_j|,\ \forall i,j=1,\dots,N\ \mathrm{with\ } i\neq j$.

\subsection*{All-to-all HTL}
We present now the A2AHTL scheme, fully described in \cite{Valerio:2016aa,VALERIO201746}, and summarised here for the reader's convenience. 
The A2AHTL solution is composed by 5 steps that are performed by each DC. We refer to Algorithm~\ref{alg:a2ahtl}.

\paragraph*{Step 0}(lines 1-5)  In the first step each DC trains a classifier on its local data. At this step there is not a unique choice for the learning algorithm to be used, it mostly depends on the specifics of the problem at hand. In this paper we use a Support Vector Machine, one of the most well known machine learning approaches. We selected it because of its good performance and simplicity for the purpose of the learning task considered in this paper. Note that if the learning task changes, the only difference would be the amount of data exchanged between nodes, while all the rest of the algorithmic details will remain the same.  After this first learning step, each DC $l_i$ holds a model $m^{(0)}_{l_i}$.

\paragraph*{Step 1}(lines 6-8) After Step 0, each DC $l_i$ sends its model $m^{(0)}_{l_i}$ to all the other DCs.  At the end of this phase each DC holds $L$ SVM classifiers, each one trained on different data.

\paragraph*{Step 2}(line 9) 
Each DC performs a second learning phase on the same local data, using the GreedyTL algorithm. In this step, the purpose of GreedyTL is to train a classifier that includes the knowledge contained in the models $m^{(0)}$ sent by all the other DCs. Precisely, GreedyTL solves an optimisation problem in order to find the linear combination of models  $m^{(0)}$ which maximises that prediction accuracy with respect to the local dataset. At the end of this step each DC has a new classifier $m^{(1)}$ that is the output of the GreedyTL algorithm.

\paragraph*{Step 3}(line 10) After the second learning phase, there is another synchronization phase where all models are sent to a unique DC, which can be any of the involved DCs.

\paragraph*{Step 4}(lines 11-14) The DC that receives all the models as per step 3 aggregates all of them into a single model $m^{(2)}$. In this paper we calculate the average model $m^{(2)} = \frac{1}{L} \sum_{l=1}^{L} m^{(1)}_{l}$. In  this paper  we can calculate the average of the models because assume that all the DCs learn the same type of model (i.e. in this case a linear classifier).  Note that depending on the specific type of models selected for the learning process it might not be possible to average the models' parameters. Therefore, in such case a different aggregation strategy needs to be adopted. 
 Moreover, once computed the final average model, all the partial models computed at step $1$ and held by each DC are no longer useful and can be removed in order to save storage. 
 
\begin{algorithm}[ht!]
	\caption{All-to-all HTL \label{alg:a2ahtl}}
	\begin{algorithmic}[1]
		\STATE Let be $L$ the number of data locations
		\STATE Let be $l_i$ the ID of the i-th location and $l_c$ the ID of the current location $c$.
		\STATE Let be $m^{(j)}$ the local model at step $j$
		\STATE Let be ${\bf X_c,y_c}$ the training patter and training labels for location $l_c$, respectively
		\STATE $m^{(0)}_{l_c} = TrainBaseLearner({\bf X_c,y_c})$ 
		\STATE SendModelToAll($m^{(0)}_{l_c}$)
		\STATE $M$= ReceiveBaseModels();
		\STATE $M \leftarrow M \cup \{m^{(0)}_{l_c}\}$
		\STATE $m^{(1)}_{l_c} = GreedyTL({\bf X_c,y_c},M)$
		\STATE SendModelToCenter($m^{(1)}_{l_c}$)
		\IF {IsCenter(c)}
		\STATE {$H$ = ReceiveGTLModels()}
		\STATE {$H \leftarrow H^{gtl} \cup \{m^{(1)}_{l_c}\}$}
		\STATE {$m^{(2)}$ = CombineModels($H$)}
		\ENDIF
		
	\end{algorithmic}
\end{algorithm}

\subsection*{Star HTL}
The StarHTL solution is a simple variation of the previous one. It is meant for those situations where the amount of data at each DC is strongly unbalanced. For example, a typical case is when the size of the local dataset at many DC is very small (e.g. even less than the size of the model to be learnt) and there is one DC that holds the majority of the data. In this case, we can save network resources by reducing the amount of models to send over the network as explained hereafter. For the sake of clarity we refer to Algorithm~\ref{alg:starhtl}

\paragraph*{Step 0}(lines 1-5) This step is unchanged w.r.t. A2AHTL. Each DC computes the model on its local data. After this first learning step, each DC $l_i$ holds a model $m^{(0)}_{l_i}$.

\paragraph*{Step 1}(lines 7-9) Election of the ``centralDC''. During this step all the DCs exchange with the others a previously selected index that will be used to elect the central node. In this paper we used, as index, the standard Information Entropy, defined as follows: $$H=-\sum_{\forall k\in K} p(k) log_{|K|}(p(k))$$ where $K$ is the set of labels and $|K|$ is the total number of labels in the learning problem at hand. Precisely, each DC computes the entropy on its local dataset and the DC with maximum entropy is the one selected to be the central DC. In this way we are selecting as centre the one with more information in its local dataset, that is, the one that is likely to train a more accurate $m^{(1)}$ model.

\paragraph*{Step2}(line 16) All the DCs ,apart from the one selected to be the centre, send their $m^{(0)}$ models to the centre DC. 

\paragraph*{Step 3}(lines 10-13) The centre DC is the only one that performs the second learning phase on its data using GreedyTL. At the end of this Step, the central DC has a new model $m^{(1)}$ built on its local data and aggregating the knowledge coming from the other models. 

\begin{algorithm}[ht!]
	\caption{Star HTL \label{alg:starhtl}}
	\begin{algorithmic}[1]
		\STATE Let be $L$ the number of data locations
		\STATE Let be $l_i$ the ID of the i-th location and $l_c$ the ID of the current location $c$.
		\STATE Let be $m^{(j)}$ the local model at step $j$
		\STATE Let be ${\bf X_c,y_c}$ the training patter and training labels for location $l_c$, respectively
		\STATE $m^{(0)}_{l_c} = TrainBaseLearner({\bf X_c,y_c})$ 
		\STATE i =ComputeCenterIndex()
		\STATE SendIndexToAll(i)
		\STATE cId=ReceiveAndSelectCenter()
		\STATE SendCenterIdToAll(cid)
		\IF {IsCenter(c)}
		\STATE $M$= ReceiveBaseModels()
		\STATE $M \leftarrow M \cup \{m^{(0)}_{l_c}\}$
		\STATE $m^{(1)}_{l_c} = GreedyTL({\bf X_c,y_c},M)$
%
		\ELSE
		\STATE SendModelToCenter($m^{(0)}_{l_c}$)
		\ENDIF
		
	\end{algorithmic}
\end{algorithm}

It is clear that StartHTL generates less traffic than A2AHTL, as (i) models $m^{(0)}$ are not sent to all  DCs, and (ii) there is no second synchronisation phase. Clearly, this might be paid in terms of lower accuracy.

%% file: results.tex
\section{Experimental setting}
\label{sec:settings}
In this section we describe the dataset we used to perform our analysis and the indexes we used to measure the performance of the system in the considered scenarios. 

\subsection{Dataset Description}
The dataset we consider in this paper is connected to the forest cover type prediction problem (CovType)\footnote{Dataset available at https://archive.ics.uci.edu/ml/datasets/Covertype}. It is a publicly available dataset that contains $581012$ observations.  Each observation is a vector of $54$ features containing both cartographic and soil information, corresponding to a $30\times30 m$ area of forest. The learning task is to use this information to predict what is the main kind of tree covering the area. The dataset contains 7 classes. Since the number of observations for each class is not the same between classes, we performed a sub-sampling of the classes in order to remove such unbalancing. Thus, the final dataset contains $19229$ points ($\approx 2700$ points per class). We used $80\%$ of the dataset as training set and remaining $20\%$ for test. 

\subsection{Performance Metrics}
In this paper we are interested in evaluating i) the energy used to transmit data or models between IoT devices and DCs, and between DCs, in order to complete the learning task, and ii) the accuracy of the model learnt through the distributed learning process.  Note that we focus only on the energy used by devices for transmitting and receiving data or models using wireless technologies.
We do not include in the energy count the energy spent by the edge server for transmitting and receiving data. Conversely we include the one used by devices for transmitting and receiving data to and from the edge server. The rational is that in this paper we are interested in analysing the impact of such system on battery powered devices. 

In this paper, we use a simplified model in which, given a certain wireless technology, the specific power for transmitting and receiving data at a fixed data rate is constant for all the nodes in the network.
We calculate the energy used for all the communications (both for transmitting and receiving data) as follows: 
\begin{equation}
    \label{eq:energy}
    E = P* t
\end{equation}
where $P$ is the power, expressed in mW, for transmitting or receiving data and $t$ is the duration, expressed in seconds, of the transmission. The duration of the transmission is calculated as follows $$t = B / S$$ where $B$ is the  uplink (or downlink) data rate expressed in bit per second (bps) and S is the number of bits to be transmitted. 

Note that $P$ and $B$ depend on the characteristics of the specific network technology. The ones considered in this papers are reported in Table~\ref{tab:nettec}. These are averages taken from reference papers found in the literature~\cite{Jensen:2012aa,Sinha:2017aa,Ahmed:2016aa,Gomez:2013aa}. 
\begin{table}
    \centering
    \begin{tabular}{|c|c|c|c|c|}
        \hline
        Wireless Tech. & Tx P.(mW) & Uplink (Mbps) & Rx P.(mW)& Downlink (Mbps) \\
        \hline
        \hline
        4G~\cite{Jensen:2012aa} & 2100 & 75  & 2100 & 35 \\
        \hline
        NB-IoT~\cite{Sinha:2017aa} & 199 & 0.2 & 199,52 & 0.2 \\
        \hline
        IEEE 802.15.4~\cite{Ahmed:2016aa} & 3 & 0.12  & 3 & 0.12 \\
        \hline
        IEEE 802.11g~\cite{Gomez:2013aa} & 1080 & 48 & 740 & 48 \\
        \hline
    \end{tabular}
    \caption{Nominal technical specifications of the considered wireless technologies}
    \label{tab:nettec}
\end{table} 

The overall energy used in the system for a session of data collection is the following: 
\begin{equation}
    \label{eq:nrg_session}
    E_S = E_C + E_L
\end{equation}
where $E_C$ is the sum of the energy used by the sensors to transmit data to the SmartMules or to the Edge server, while $E_L$ is the energy used by the SmartMules during the distributed learning process. 

In this paper, in order to evaluate the prediction performance  of our system we use the well known F-measure~\cite{Powers:2011aa}.  The F-measure is a common aggregate index mostly used to evaluate the performance of a content retrieval system or a classifier. We use it because it provides a more precise description of the performance of our system with respect to the simple accuracy index defined as the number of correct predictions divided by the total number of predictions made. Precisely, through the F-measure we are able to summarize in one single index the performance expressed by the Precision and Recall indexes. The F-measure is defined as the harmonic mean between the precision index and the recall index, hereafter defined. Note that in this paper all the performance indexes are location-wise, i.e. they always refer to performance obtained by models trained in each separate location $l$.

The precision (or specificity) index is  defined as the number of correct predictions divided by the total number of predictions made. More formally: 
\begin{equation}
    p_l =  \frac{1}{n_l}\sum_{i=1}^{n_l} I(y_i,\hat{y_i})
\end{equation}
with 
$$I(y_i,\hat{y_i})=\left\{\begin{array}{c c} 1 & \mathrm{if\ } y_i=\hat{y_i} \\ 0 & \mathrm{otherwise}\end{array}\right. $$ where $\hat{y}$ is the predicted class of the i-th pattern $x_i$ and $y_i$ is its true class and $n$ is the number of elements in the test set.  

The recall (or sensitivity) index is defined as the number of correct predictions for a specific class divided by the number of the patterns that belong to that specific class, averaged over all classes. More formally: 
\begin{equation}
    r_l = 
    \frac{1}{|C|}\sum_{\forall c \in C}\frac{1}{n_{l,c}}\sum_{i=1}^{n_{l,c}}I(y_{c,i},\hat{y}_{c,i})
\end{equation}
Finally the F-measure is defined as follows:
\begin{equation}
    F_l = 2*\frac{p_l \cdot r_l}{p_l+r_l}
\end{equation}
This measure takes values in the range $[0,1]$, where $0$ stands for the worst prediction performance and $1$ to the best one.

\section{Results}
\label{sec:results}
We present now the results of our analysis. We recall that our interest is to identify through simulation how to configure the learning process such that, given a minimum target loss of accuracy (with respect to a centralised solution), the total amount of energy used to transfer data to the devices involved in the learning process is minimised. 
All the results we present hereafter are average values over 10 simulations. Confidence intervals, which we computed with a 95\% confidence level, are very tight, therefore we do not show them, to  improve the clarity of the plots.

Before presenting the results we provide the details that are shared by all the scenarios we analysed in this paper. As we explained in Section \ref{sec:statement}, we are considering a process of data collection through time. The process is divided in 100 time windows. In our simulations in each window are collected 100 observations, each one containing 54 features (the number of features is dataset dependent). At the end of each collection window, a learning phase takes place. The learning phase can be either distributed or centralised, depending on where the data has been sent. If all the data has been sent to the edge server, the learning phase will be centralised, while if the data is partially on the edge and partially (or totally) on the mules, the learning phase will be done using an HTL approach.
Remember that regarding the last two cases, the learning algorithm is iterative and this is reasonable, as those nodes are sensing devices with limited storage, which typically have to flush their storage after a while. We remark that, although the specific dataset we use for evaluation the dynamic over time is slowly varying, we might expect that in cases where such a dynamic is quicker and data sensed in different time windows could not be that correlated, the learning performance would be different. However, we think that the results shown in the paper for the initial time slots provide indications of the performance obtained in these cases, where limited or no prior knowledge can be carried over between time slots.
Remember that according to the scenario described in Section~\ref{sec:statement}, in each collection interval, the number of mules that have collected data is a realisation of a random variable distributed according to a Poisson distribution of rate $\lambda$, while the number of data on each of such mules is distributed according to a Zipf law of parameter $\alpha$. Specifically, we consider $\lambda=7$ and $\alpha=1.5$. We chose these parameters in order to simulate a scenario in which the data collection process is highly unbalanced, i.e. few SMs hold the vast majority of the data.


\subsection{Benchmark Scenario: All data on the Edge Server}
\begin{figure}[ht]
    \centering
    \subfloat[]{
        \includegraphics[width=.5\textwidth]{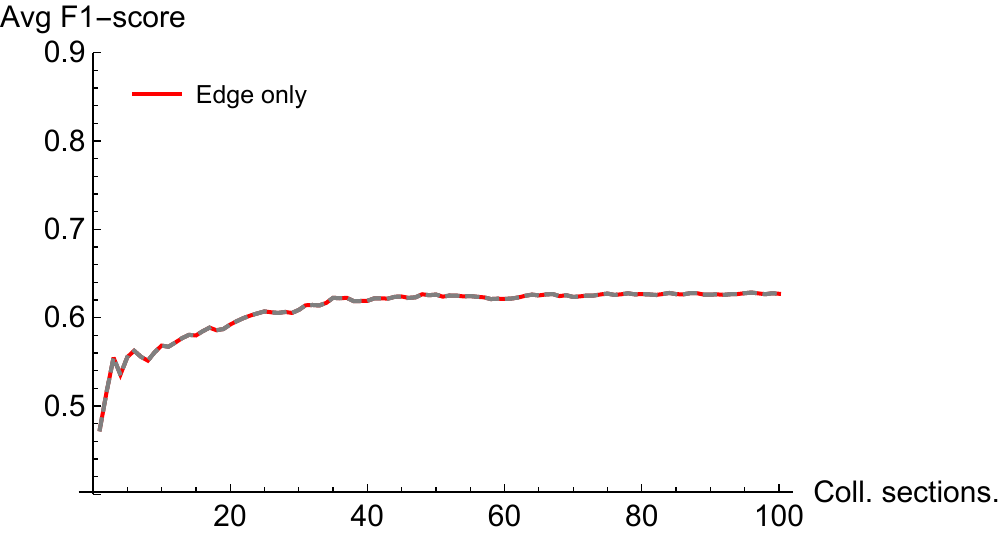}
    }
    \subfloat[]{
        \includegraphics[width=.5\textwidth]{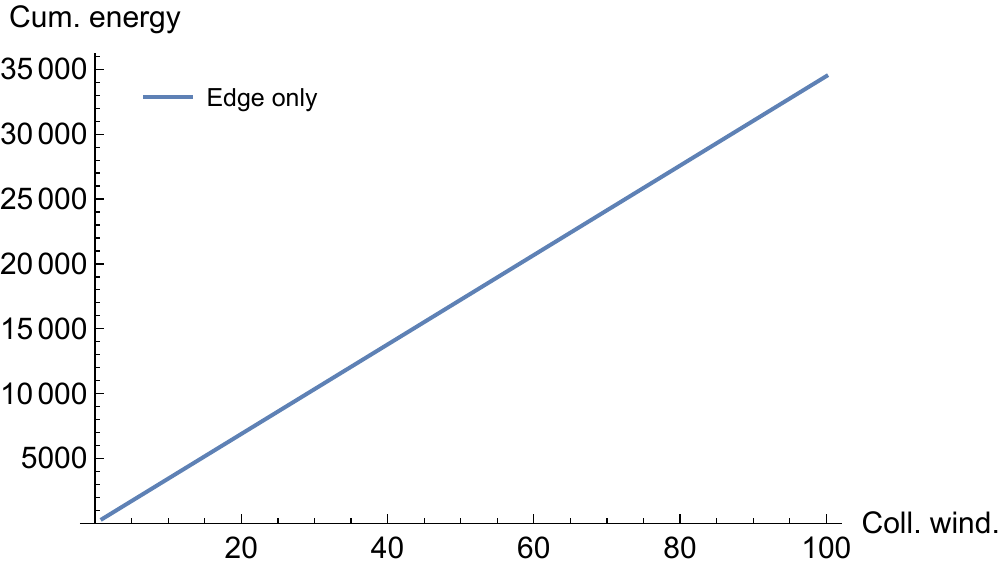}
    }
    \caption{Accuracy and cumulative energy cost for the edge only solution using NB-IoT}
    \label{fig:edgeonly}
\end{figure}

Let us now present the analysis of the scenario that we use as benchmark during the rest of the paper. Precisely our benchmark is a scenario in which there are no mules in the area during the entire collection process. Therefore, for each collection session, all the data is sent to the edge server through NB-IoT, a cellular technology specifically designed for IoT devices. On the edge server, for each collection window the model is updated according to the new data. Results are shown in Figure~\ref{fig:edgeonly}. In terms of accuracy the final F1-value that we obtain using a linear classification model is $0.63$\footnote{This final value corresponds to a centralised training session on the entire dataset.}. In absolute terms is not a particularly high value but is the best we can obtain from a linear model applied to this problem. However in this paper we are not interested in finding the best model that fits  the data, but we are interested in analysing if through a distributed learning approach based on HTL it is possible to save resources while obtaining performance comparable to the one obtained executing the learning in a centralised fashion.
Regarding the energy consumed for transmitting all the data using a NB-IoT we found a final value of $34477$ mJ that, for the rest of the paper we will use as benchmark in all the considered scenarios. Note that the values of accuracy and energy reported at the end of the collection process are equivalent to those obtained by the best performing centralised algorithm (in this case SVM) having access to the complete dataset. 

\subsection{Scenario 1: Partial data on Edge}
In this scenario we want to evaluate if it is possible to save resources limiting the amount of data that is transmitted to the edge. To this end, we analysed tree separate cases in which we progressively decrease the amount of data sent to the edge server. Specifically we are simulating a scenario in which the SMs present in the area are able to collect only a portion of all the data produced by the sensors and the remaining one must be transmitted to edge in order to not be wasted. In this way, we are simulating a situation where no SMs are in proximity of some of the IoT nodes, thus the latter have to transmit their data to the Edge Server. In this scenario the IoT sensors use  IEEE 802.15.4 for transmitting data to mules and if no mules are present, they use NB-IoT for sending data to the ES. Regarding the learning phase, in this scenario we present the performance accuracy obtained with StarHTL. We discuss only this method because in this specific scenario both of them obtain the same accuracy and StarHTL is  lighter in terms of energy consumption.  In this scenario, the SMs that are involved in the learning process communicate with each other and with the ES using 4G technology, i.e.,  we are allowing long range communications between the SMs. 
\begin{figure}[ht]
    \centering
    \subfloat[]{
        \includegraphics[width=.5\textwidth]{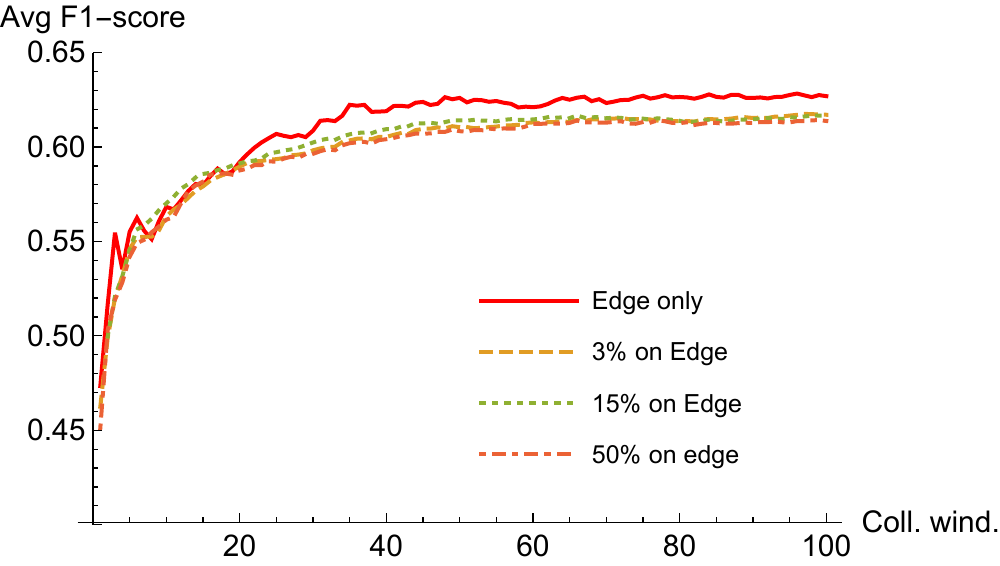}
    }
    \subfloat[]{
        \includegraphics[width=.5\textwidth]{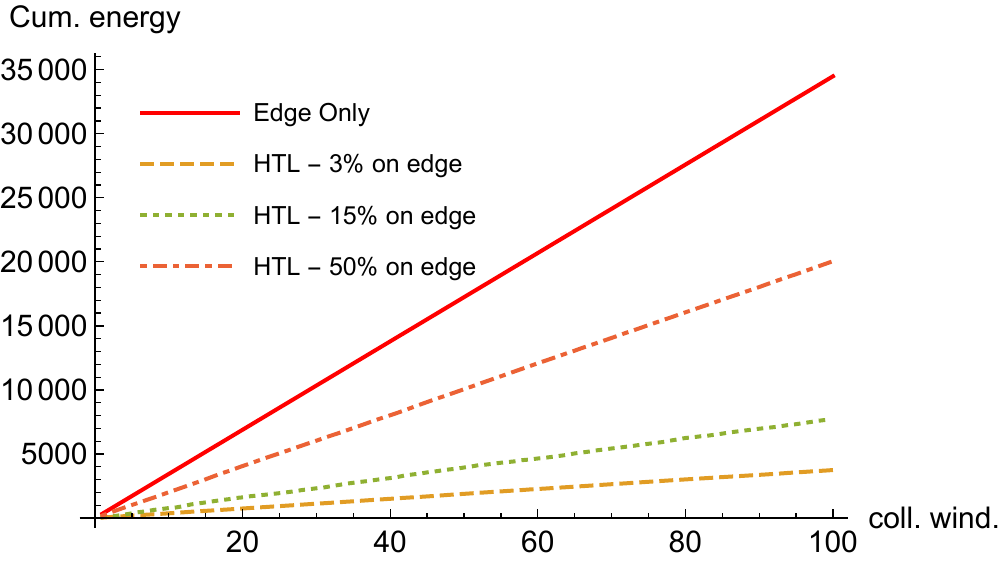}
    }
    \caption{Accuracy and cumulative energy cost for different percentages of data on edge.}
    \label{fig:withedge}
\end{figure}

\begin{figure}[ht]
    \centering
    \subfloat[$3\%$ Edge]{
        \includegraphics[width=.3\textwidth]{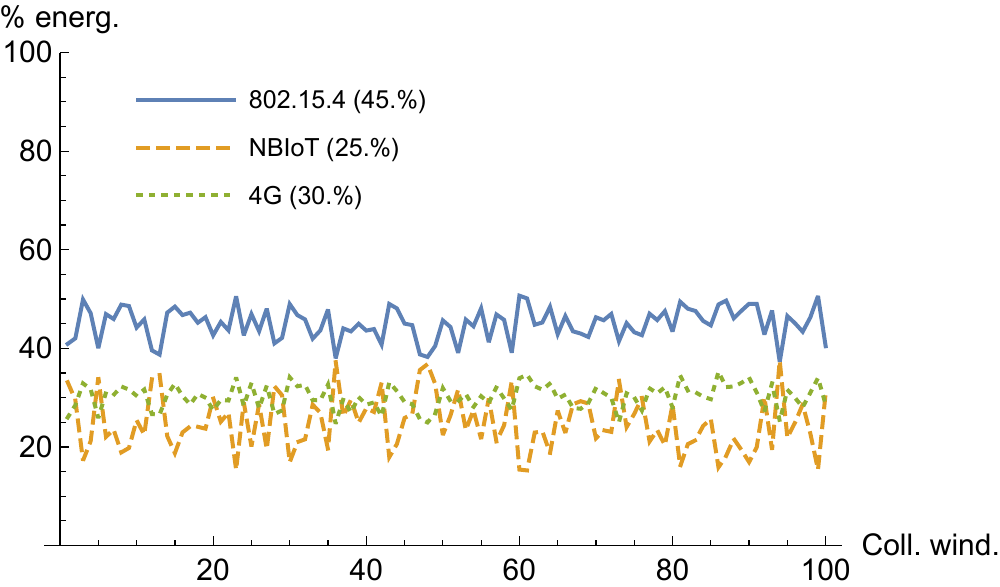}
    }
    \subfloat[$15\%$ Edge]{
        \includegraphics[width=.3\textwidth]{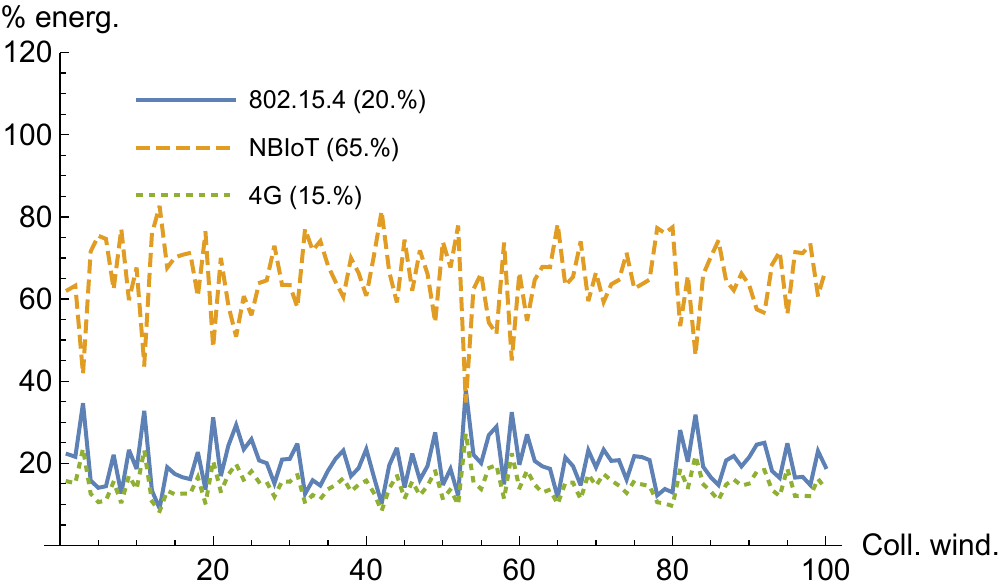}
    }
    \subfloat[$50\%$ Edge]{
        \includegraphics[width=.3\textwidth]{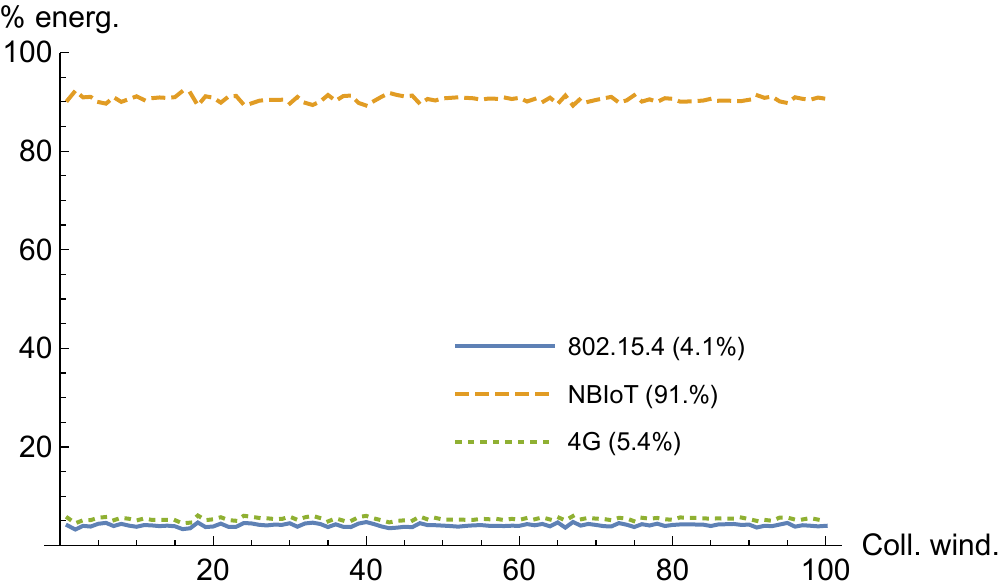}
    }
    \caption{Composition of the energy consumption for the tree scenarios. }
    \label{fig:withedgeperc}
\end{figure}

As far as the energy consumption in concerned, we can see from Figure~\ref{fig:withedge} that collecting data through mules and sending the remaining data to the edge has the effect of saving energy resources. The reason of this is the fact that the  transmission power of NB-IoT is greater than the one of 802.15.4, i.e. the former is 199 mW while the latter is 3mW. This makes NB-IoT more expensive than 802.15.4. 
It might look surprising that a technology like NB-IoT, which is designed to be energy efficient, turns out to be so energy hungry in our case. However, this is due to the fact that NB-IoT is designed having in mind short and infrequent transmissions. Our results show that it  might not address IoT applications in general, as in cases where significant amount of data have to be transferred from IoT devices, it becomes energy hungry (at least, with respect to short-range technologies).
Therefore, introducing a short/medium range technology such as 802.15.4 for data collection is beneficial for the entire system. 
This is even more evident looking at Figure~\ref{fig:withedgeperc}b-c, which shows that, when the amount of data sent through NB-IoT becomes significant, the great part of the energy consumed for transmission is due to the data collection process. Conversely, the energy spent for the learning process and the short range data collection are comparable. 
However, looking at Figure~\ref{fig:withedgeperc} we see that the data collection (which is done through IEEE 802.15.4 and NB-IoT) is more energy expensive than then the learning process (which is done through 4G), although, in absolute terms 4G is far more expensive that IEEE 802.15.4. The main reason for this behaviour is that StarHTL generates a very limited amount of traffic, proving it to be a very efficient and effective distributed learning  solution. In fact in every scenario StarHTL is able to obtain an accuracy performance very close to the centralised one, i.e. the error with respect to the centralised solution is $\lesssim 2\%$. Note that this value is computed on the F1-score values obtained in the interval from the 50th to the 100th collection window, i.e. this interval is the one where the learning process converges.
 
In table \ref{tab:withedgegain} we report the values of the energy consumed and the corresponding gain with respect to the Edge-only configuration, when a different amount of data is collected by mules instead of the edge server. Reducing the amount of data transmitted provides a saving of energy up to $89\%$.
\begin{table}[ht]
    \centering
    \begin{tabular}{|c|c|c|c|c|}
    	\hline
        Scenario & Energy (mJ) & Gain (\%) & F-Measure & Acc. Loss \\
        \hline
        Only Edge & 34477 & 0 & 0.63 & 0\% \\
        50\% on Edge Server & 20047 & 42 \% & $0.61$ & 2\%\\
        15\% on Edge Server & 7736 & 77 \% &$0.61$ & 2\% \\
        3\% on Edge Server& 3749 &  89 \% &$0.61$ & 2\%\\
        \hline
    \end{tabular}
    \caption{Energy consumption and accuracy with different amount of data collected on the mules}
    \label{tab:withedgegain}
\end{table}

\subsection{Scenario 2: No data on edge server}
\label{ssec:scenario2}
From the previous results we found that not only it is more convenient from the energy consumption point of view limiting the amount of data to transmit on the edge server but it also allows to obtain learning performance comparable with a centralised  solution. 
In this section we want to further stress this idea, completely removing the collection of data on the edge and assuming that all the data is collected by the SMs. The scenario at hand is the very same of the one considered before, i.e. the SMs receive through 802.15.4 all the data coming from the IoT sensors and after each collection window a learning session is performed. In this section, we compare the performance of the two learning mechanisms presented in Section \ref{sec:methodology}, namely A2AHTL and StarHTL.

\begin{figure}[ht]
    \centering
    \subfloat[F1]{
        \includegraphics[width=.5\textwidth]{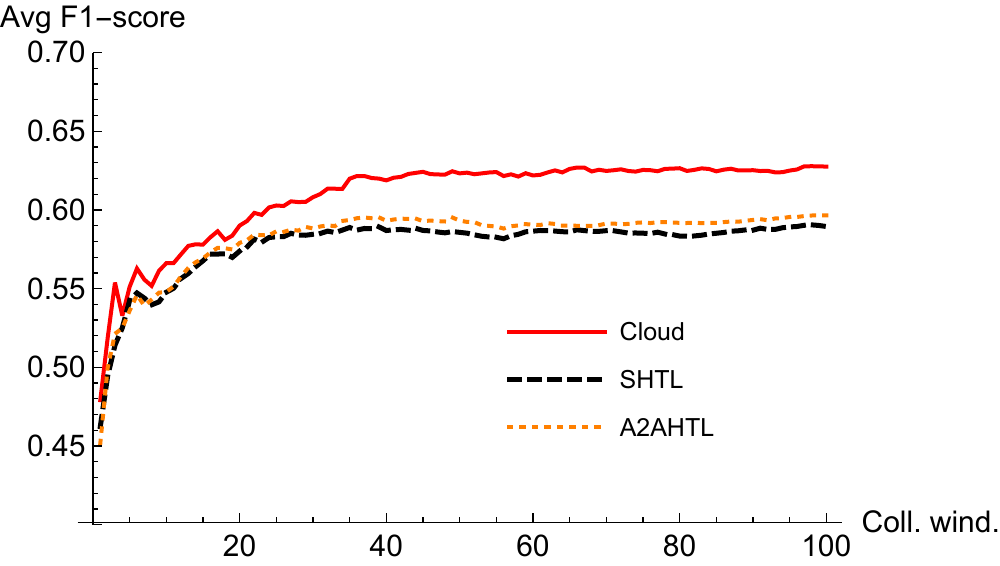}
    }
    \subfloat[Err ]{
        \includegraphics[width=.5\textwidth]{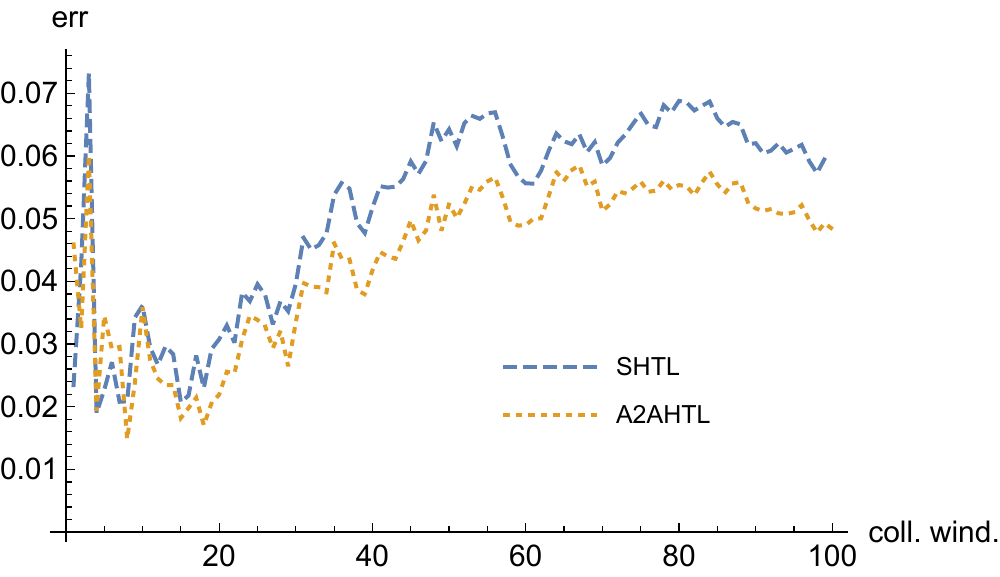}
    }
    \caption{Accuracy  for A2AHTL and SHTL compared to the EdgeOnly solution, when no data is sent to the Edge. Initial data distribution is Zipf.}
    \label{fig:accnoedge}
\end{figure}

Looking at Figure~\ref{fig:accnoedge} we note that in this case, the loss of accuracy of both the HTL approaches passes from  $\sim 2\%$ of  error to the $\sim 5\%$ for A2AHTL and $\sim 6\%$ for SHTL. This can be explained considering how data is distributed across the mules. We recall that according to the Zipf distribution, there is one mule with the vast majority of data (in this case one mule holds $55\%$ of the data) while the others have the rest. Therefore in this scenario, although the number of mules in the area is on average $7$, many of them hold just a very little portion of the existing data for each collection window. The immediate effect of this strong unbalancing is that the mules with few data observation will build inaccurate models and, since they are the majority, they do not positively contribute to the overall learning process. This is specially true for the SHTL approach, where only the mule designated as centre is the one that performs the retraining using the models from the other mules. In A2AHTL instead, the retraining is done by all the SMs, meaning that each of them uses the knowledge of the others to build the second model, and this has the effect of mitigating the problem arisen with SHTL. 

Let us now look at Table~\ref{tab:mule} that contains the values of energy used by the collection  and the learning processes. Regarding the latter, we compare two wireless technologies, i.e. 4G and IEEE 802.11g. Our aim is to investigate if the usage of short range wireless technology contributes to a further reduction of the energy consumption in comparison with the usage of the 4G technology. We used IEEE 802.11g instead of the newer version  of the protocol (i.e. IEEE 802.11n) because it is still the most diffused one. Before commenting the results it is important to specify that changing from 4G to Wifi, imposes us to modify the communication scheme between the SMs. In fact, we assume a star topology where one of the SMs acts as Access Point, initializing the wireless network to which all the other mules will connect and use to communicate. This is the configuration, for example, of WiFi Direct, which needs one device to act as Access Point for the others~\cite{Arnaboldi:2017aa,Conti:2013aa}.

From the energy consumption point of view, looking at the first two rows (4G case) of Table \ref{tab:mule} we see that using only the mules instead of using a combination of edge and mules for data collection and processing, further limits the costs. In fact, looking at the energy used by the two approaches we notice that SHTL is a more preferable choice the it is more energy efficient, i.e. we save $93\%$ of energy w.r.t. the Edge Only solution, and the loss in accuracy with respect to A2AHTL is contained. In fact, compared to A2AHTL, SHTL is just $1\%$ less accurate with respect to the centralised solution.
\begin{table}[ht]
    \centering
    \begin{tabular}{|c|c|c|c|c|}
    	\hline
        Algo & Cum. En.(mJ) & Data Coll. (mJ) & Learning En. (mJ) & Gain (\%)  \\
        \hline
        A2AHTL - 4G & 2789 & 1728 & 1061 & 91\%\\
        SHTL - 4G & \bf2513 & 1728 & 785  & \bf93\%\\
        A2AHTL - wifi & 5184 & 1728 & 3456 & 89\% \\
        SHTL - wifi & \bf2066 & 1728 & 338 & \bf94\%\\
        \hline
    \end{tabular}
    \caption{Performance when all data is collected by the SMs.}
    \label{tab:mule}
\end{table}

Considering the WiFi case, looking at the third and fourth row of Table \ref{tab:mule} we notice the benefit of using IEEE 802.11g in combination with SHTL. In fact in the A2AHTL case, we notice that the energy spent for the learning process increases w.r.t. the one used by 4G. We justify this result by recalling that although IEEE 802.11g uses less transmission power, it has half of the downlink data rate of 4G. Moreover with the star topology managed directly by the mules, the number of transmissions made by battery powered devices increases. Conversely, for SHTL is even more convenient using IEEE 802.11g instead of 4G. The reason is that SHTL limits the number of models to be sent by the mules, i.e. we recall that in A2AHTL each model is sent to all the others while in SHTL only to the mule acting as  centre. In light of this results we can state that in these settings if the mules have the possibility to communicate with each other using a short/medium range wireless technology such as IEEE 802.11g, it is possible to save up to $94\%$ of energy with respect to the Edge Only solution, with a loss in accuracy in the range of $6\%$. Interestingly, the configuration of the HTL framework may play a significant role: if the loss of accuracy of SHTL with WiFi (the most energy efficient configuration) is not acceptable, then it is better to switch to 4G for communication between the mules, as A2AHTL with WiFi is less energy efficient than both HTL versions with 4G.

 One aspect that affected the results obtained in the scenarios considered so far is that some of the SMs collected too few observations. Therefore, for these specific nodes, it would be more energy efficient to transmit data (which are few) instead of the local model.     
  To cope with this problem, we adopt a        very simple heuristic that is based on the idea that it is better to send models only when the amount of local data is above a given threshold. In our case the threshold is set to twice the size of the model. The mules that have an amount of local data below the threshold,  aggregate their data  on one of them until the condition is satisfied. Only the node that receives the data takes part to the learning process.
In this specific case, after having applied this rule, the number of nodes involved in the distributed learning process passes from 7 to 3.

\begin{figure}[ht]
	\centering
	\subfloat[F1]{
		\includegraphics[width=.5\textwidth]{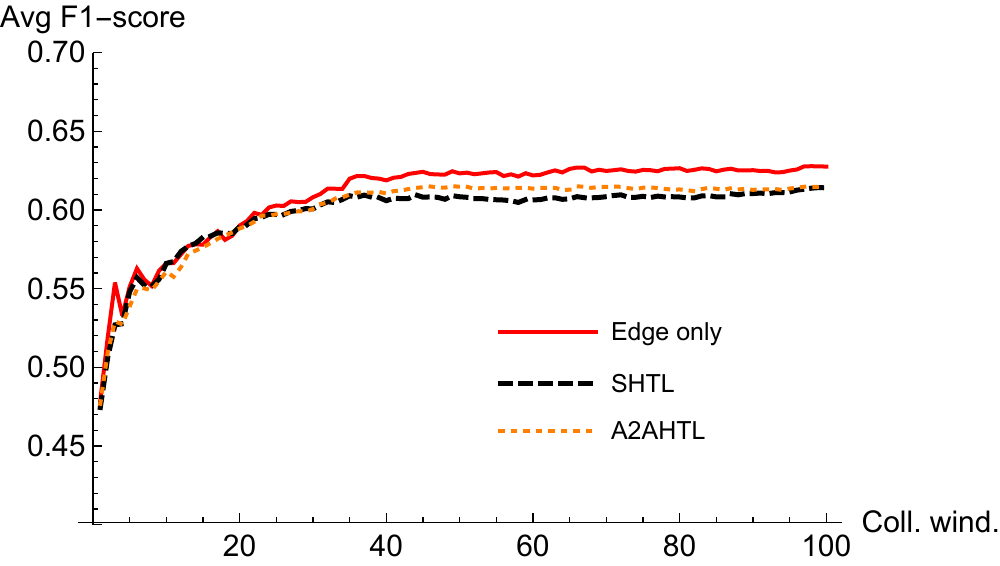}
	}
	\subfloat[Err]{
		\includegraphics[width=.5\textwidth]{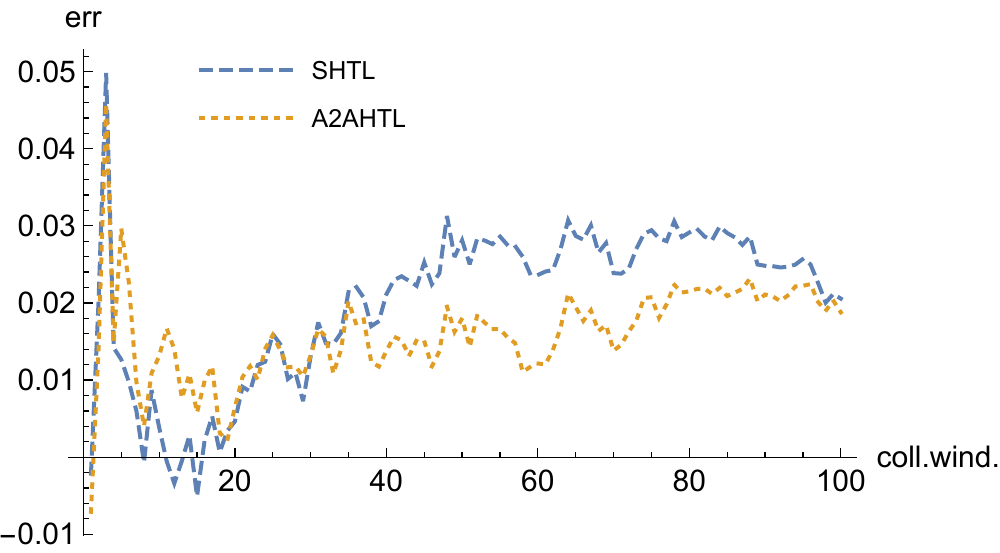}
	}
	\caption{Accuracy  for A2AHTL and SHTL with "data aggregation heuristic" compared to EdgeOnly solution. Zipf data distribution.}
	\label{fig:accagrnoedge}
\end{figure}

Our heuristic is very simple, but it proves to be effective in terms of accuracy. In fact, looking at Figure \ref{fig:accagrnoedge} we see that the  accuracy is improves with respect to the HTL configuration without aggregation. Specifically, SHTL is 3\% less accurate than the centralised solution while A2AHTL is only 2\% less accurate. It is worth noting that we obtained the very same loss of accuracy obtained in the \emph{Scenario 1} without accumulating data on the Edge Server. Moreover, regarding the energy used to obtain such results we notice that, SHTL is always more energy efficient and also in this case, using WiFi instead of 4G proves to be beneficial, i.e. 94\% of energy saved using WiFi and 93\% using 4G. Also, note that in this case we avoid the increase of energy consumption of A2AHTL with WiFi, thus making using WiFi always more energy efficient than using 4G.

\begin{table}[ht]
    \centering
    \begin{tabular}{|c|c|c|c|c|}
    	\hline
        Algo & Cum. En.(mJ)& Coll. En. (mJ)& Learn. En. (mJ) & Gain (\%) \\
        \hline
        A2AHTL - 4G & 2957 & 1728 & 1229 & 91\%  \\
        SHTL - 4G & 2849 & 1728 & 1120 & 92\%\\
        A2AHTL - wifi & 2700 & 1728 & 1028 & 92\% \\
        SHTL - wifi & 2211 & 1728 & 483 &94 \%\\
        \hline
    \end{tabular}
    \caption{Performance when all data is collected by the SMs with data aggregation}
    \label{tab:muleaggr}
\end{table}

\subsection{Scenario 3: No data on the Edge Server. Uniform initial data distribution.}
\label{ssec:scenario3}
\begin{figure}[ht]
	\centering
	\subfloat[F1]{
		\includegraphics[width=.5\textwidth]{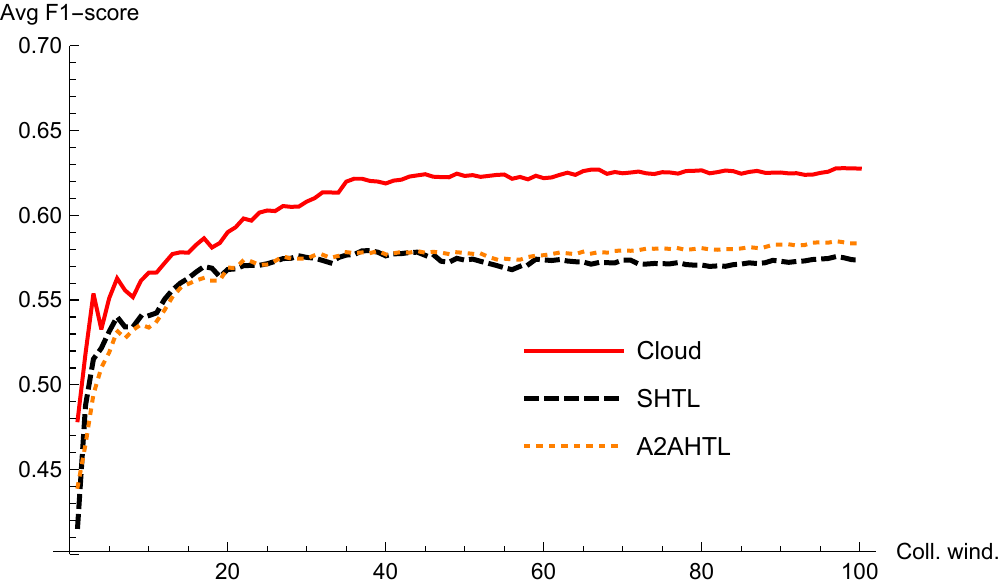}
	}
	\subfloat[Err]{
		\includegraphics[width=.5\textwidth]{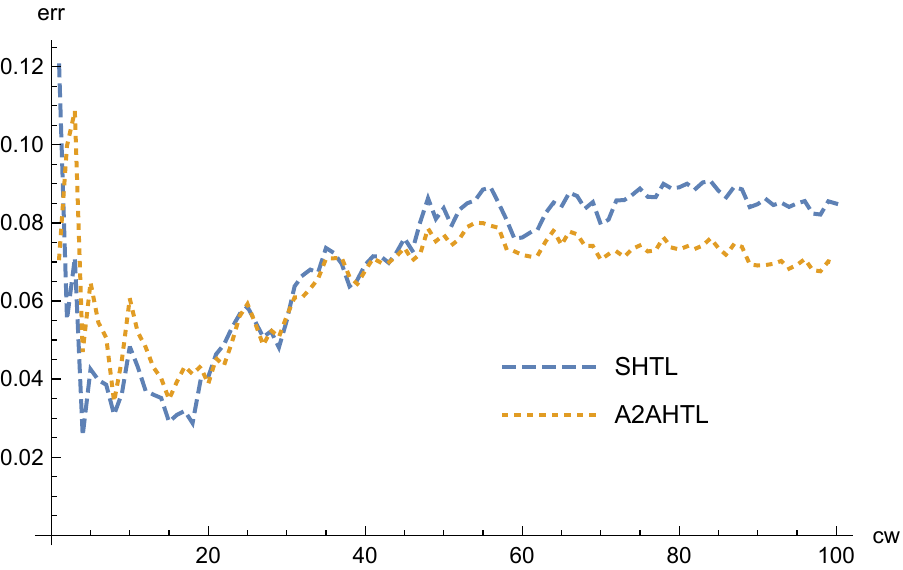}
	}
	\caption{Accuracy  for A2AHTL and SHTL compared to the EdgeOnly solution. Uniform distribution.}
	\label{fig:accnoedgeu}
\end{figure}

 So far we presented results considering that the distribution of data collected by SMs is strongly unbalanced. Now we investigate if the conclusions drawn for the Scenario 2 are still valid when we impose a different initial distribution of data between the mules. To this end, we consider a data collection scenario where each mule collects almost the same amount of data. Precisely, the amount of data that each mule collects, in this scenario is uniformly distributed, i.e. the amount of data per mule is in the range of $14\%$. Note that the main difference with the previous scenario is that in this case there is not a single mule with more data than the others. 
Here, we performed the very same analysis done in Scenarios 1 and 2. In Figure~\ref{fig:accnoedgeu} the accuracy performance of both A2AHTL and SHTL, without data aggregation, are shown. Differently from the accuracy results presented in Scenario 2, we notice a degradation of performance, i.e. the loss in accuracy w.r.t. the centralised solution is in the range of $7\%$ and $8\%$ for A2AHTL and SHTL, respectively. This result is mainly due to the amount of data present at each SM. In fact since each of the SM has a very limited amount of local data (i.e. only $14$ observations each) the initial models build at \emph{Step0} of both A2AHTL and SHTL procedure under-fit the data, i.e. the models are not enough representative of the data on which they are built. Therefore the combination of these models results in a loosely accurate one. Instead, looking in Table~\ref{tab:muleu} we notice the very same behaviour observed in Scenario 2, that is, SHTL is more energy efficient than A2AHTL and the wireless technology that allows to save more resources, between the ones considered in this paper, is IEEE 802.11g. 

\begin{table}[ht]
	\centering
	\begin{tabular}{|c|c|c|c|c|}
		\hline
		Algo & Cum. En.(mJ) & Data Coll. (mJ) & Learning En. (mJ) & Gain (\%)  \\
		\hline
		A2AHTL - 4G & 3350 & 1728 & 1622 & 90\%\\
		SHTL - 4G & 2753 & 1728 & 1025  & \bf92\%\\
		A2AHTL - wifi & 6676 & 1728 & 4948 & 80\% \\
		SHTL - wifi & 2169 & 1728 & 441 & \bf94\%\\
		\hline
	\end{tabular}
	\caption{Performance when all data is collected by the SMs. Uniform distribution.}
	\label{tab:muleua}
\end{table}

\begin{figure}[ht]
	\centering
	\subfloat[F1]{
			\includegraphics[width=.5\textwidth]{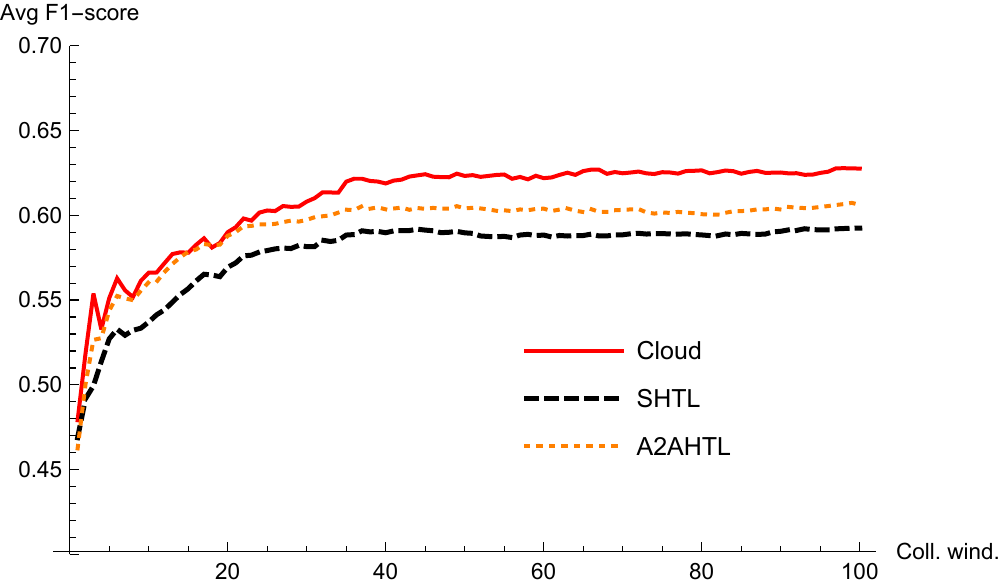}
	}
	\subfloat[Err ]{
		\includegraphics[width=.5\textwidth]{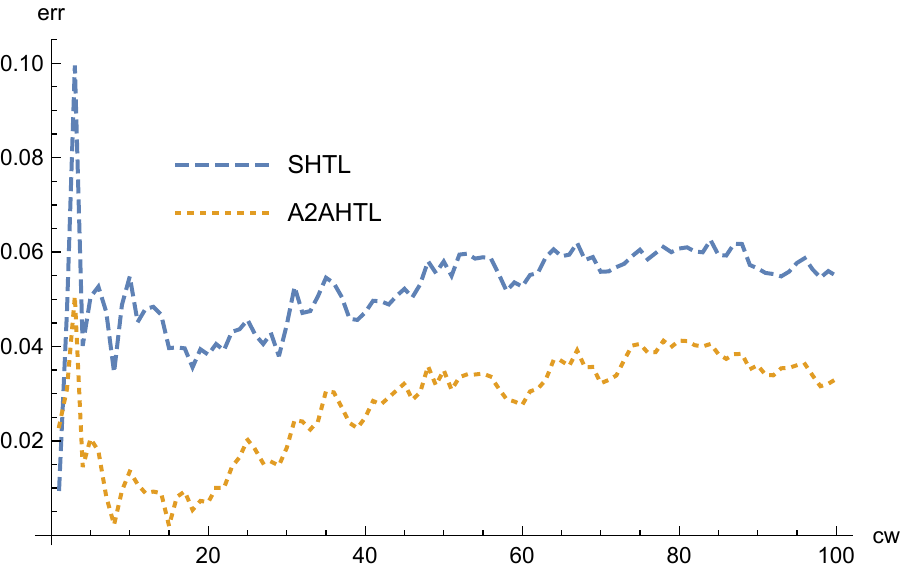}
	}
	\caption{Accuracy for A2AHTL and SHTL with "data aggregation heuristic" compared to EdgeOnly solution. Uniform initial data distribution.}
	\label{fig:accagrnoedgeu}
\end{figure}

In order to evaluate if it is possible to reduce the  loss of accuracy observed in the previous set of results, we apply also in this scenario the data aggregation heuristic, in order to increment the amount of data used by each mule to execute the distributed learning process. In this case, after the data aggregation process, the average number of nodes involved in the learning process passes from 7 to 3. Looking at Figure~\ref{fig:accagrnoedgeu}, which shows the accuracy for both learning approaches, we see that SHTL obtains the best performance, reducing the loss of accuracy to $3\%$ (see Table~\ref{tab:muleua}). Also in this final case, the data aggregation heuristic proves to be beneficial for improving the accuracy but, like in the previous case, at a marginal increment in terms of energy cost. 

\begin{table}[ht]
	\centering
	\begin{tabular}{|c|c|c|c|c|}
		\hline
		Algo & Cum. En.(mJ) & Data Coll. (mJ) & Learning En. (mJ) & Gain (\%)  \\
		\hline
		A2AHTL - 4G & 3517 & 1728 & 1789 & 90\%\\
		SHTL - 4G & 3235 & 1728 & 1507  & \bf91\%\\
		A2AHTL - wifi & 3885 & 1728 & 2157 & 89\% \\
		SHTL - wifi & 2377 & 1728 & 649 & \bf93\%\\
		\hline
	\end{tabular}
	\caption{Performance when all data is collected by the SMs.  Uniform distribution with data aggregation.}
	\label{tab:muleu}
\end{table}

%% file: complexity.tex
\section{Computational Complexity}
\label{sec:complex}

Training a model could be computationally intensive and this might strongly affect the resource budget of battery powered devices. 
Specifically, the energy consumption related to computation depends primarily on the computational complexity of the part of the learning algorithm executed by mules. It would also depend on the specifics of the hardware and processing architecture of the mules, which would be dependent on the particular nodes used.
Therefore, it is interesting to analyse the computational impact that our distributed learning procedure might have on such kind of devices.
To make the analysis more general, we have (i) analysed the complexity of the algorithm implemented at the mules, and (ii) analysed the loss of accuracy when we reduce this complexity. Specifically, the complexity of the algorithm executed by mules, which is a function of the number of data points used for re-training the local models. We then analyse, via simulation, the impact on accuracy of reducing the number of data points. We show that the loss of accuracy is very little, even in case of a drastic reduction of the number of data points used. This shows that it is possible to significantly limit the energy spent in computation without loosing significantly in terms of accuracy.

In our solution the two main computational demanding blocks are i) the training of the base model (in our case a linear SVM) and ii) the re-training using the GreedyTL algorithm. As shown in \cite{Shalev-Shwartz:2008aa} the computational complexity to train a linear SVM is $O(N^2)$, where $N$ is the size of the dataset. In our scenario we recall that the i-th SM   holds a portion of size $n_i$ of the entire dataset, i.e., $N=\sum_{i=1}^m n_i$, where $m$ is the total number of SMs, thus the computational effort, related to the SVM step, of  the i-th device is $O(n_i^2)$.
This means that system-wise the computational complexity is $O(\sum_{i=1}^m n_i^2)$. This is by far less than $O(N^2)$, because it holds that: $\sum_{\forall i}n_i^2 < (\sum_{\forall i}n_i)^2$.
Regarding GTL, for a single device, the complexity for training a model includes i) the complexity for executing the $m-1$ source models, which in case of a linear model is $O(n)$ and ii) the complexity for building the model which is, according to \cite{Kuzborskij:2017aa} $O(n^2)$. Therefore, the overall device-wise complexity is quadratic on the number of local data held by a device and, system-wise, it is $O(\sum_{i=1}^m n_i^2)$.

It is important to note that in our solution the SVM training can be considered as part of the problem definition, i.e., a design choice imposed to satisfy a constraint of the system. This means that, depending on the type of scenario, it could be replaced by another learning solution, with different computational requirements. Conversely, we analyse how the computational complexity of our solution changes when varying the configuration of GTL.
Interestingly, as also shown in \cite{OrabonaGreedyTL}, one key feature of GreedyTL is that it proves to be very effective even when the amount of local data is small. This, can be exploited to significantly reduce the computational burden needed for training the model. For example, in \cite{OrabonaGreedyTL,Kuzborskij:2017aa} is shown that less than 10 data points per class are enough to train successfully a model. 
Therefore, as we show in the rest of this section, it is possible to control the impact of GreedyTL on the overall computational effort at a small cost in terms of accuracy.

To this end, we performed a set of simulations in which we incrementally decrease the number of data points per class provided to GTL. Precisely, we are interested in the relation between the number of data points used for the training (i.e., the operating computational complexity) and the accuracy obtained by the learning process. 
	
In the following we present the results for all the cases considered in the paper, namely StarHTL and A2AHTL applied to Scenario 2 and 3. In all the considered scenarios, for each collection window there are 7 SMs (the same value used in the rest of the paper) and we evaluated both StarHTL and A2AHTL in which GreedyTL is trained using a random sample of size 2,5,10 data points, for each class. This quantities in the case of Scenario 3 (i.e., uniform data distribution) are equivalent to using, on average, the $15\%, 40\%, 75\%$ of the data collected by each node, respectively. Conversely, since in Scenario 2 the data distribution is Zipf, such quantities are different from SM to SM. For the sake of clarity we reported in Table \ref{tab:datadist} the average percentages of local data used by each device during training. Note that SM1 is the device with more data and SM7 is the one with less data for each collection window.

\begin{table}
	\centering
	\begin{tabular}{|c|c|c|c|c|c|c|c|}
		\hline 
		Abs. Sample Size & SM1 & SM2 & SM3 & SM4 & SM5 & SM6 & SM7 \\ 
		\hline 
		2 & 4\% & 9\% & 18\% & 33\% & 40\% & 50\% & 100\% \\ 
		\hline 
		5 & 10\% & 22\% & 45\% & 83\% & 100\% & 100\% & 100\% \\ 
		\hline 
		10 & 19\% & 45\% & 90\% & 100\% & 100\% & 100\% & 100\% \\ 
		\hline 
	\end{tabular} 
\caption{Percentage of local data used by each device corresponding to the sub-sample sizes for Scenario 2.}
\label{tab:datadist}
\end{table}

\begin{figure}[ht!]
	\centering
	\subfloat[A2AHTL]{
	\includegraphics[width=.45\textwidth]{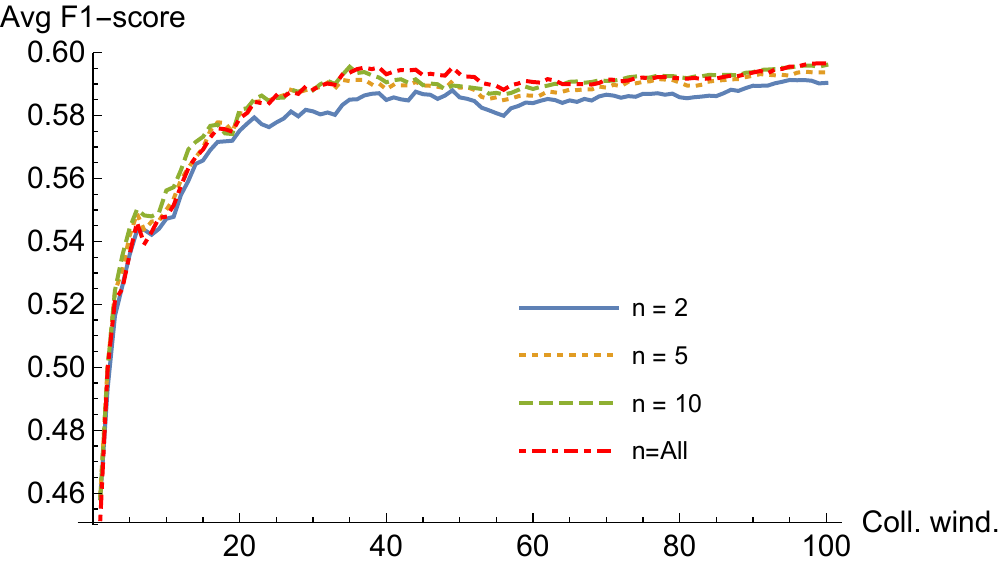}
	}
\subfloat[SHTL]{
	\includegraphics[width=.45\textwidth]{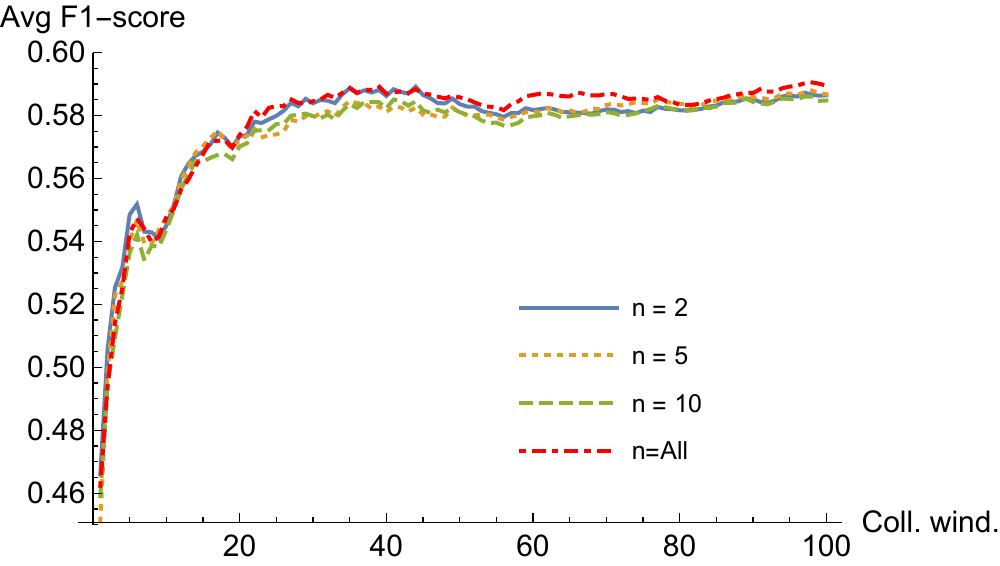}}
	\caption{Accuracy for different sizes of sample used by the re-training part. Zipf distribution of data on SMs.}
	\label{fig:complacc_zipf}
\end{figure}

\begin{figure}[ht!]
	\centering
	\subfloat[A2AHTL]{
		\includegraphics[width=.45\textwidth]{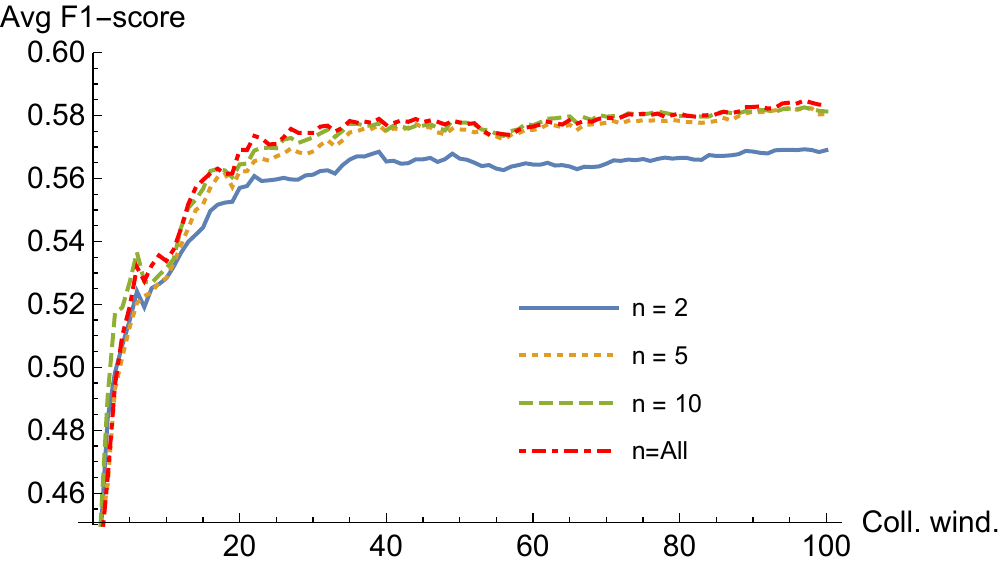}
	}
	\subfloat[SHTL]{
		\includegraphics[width=.45\textwidth]{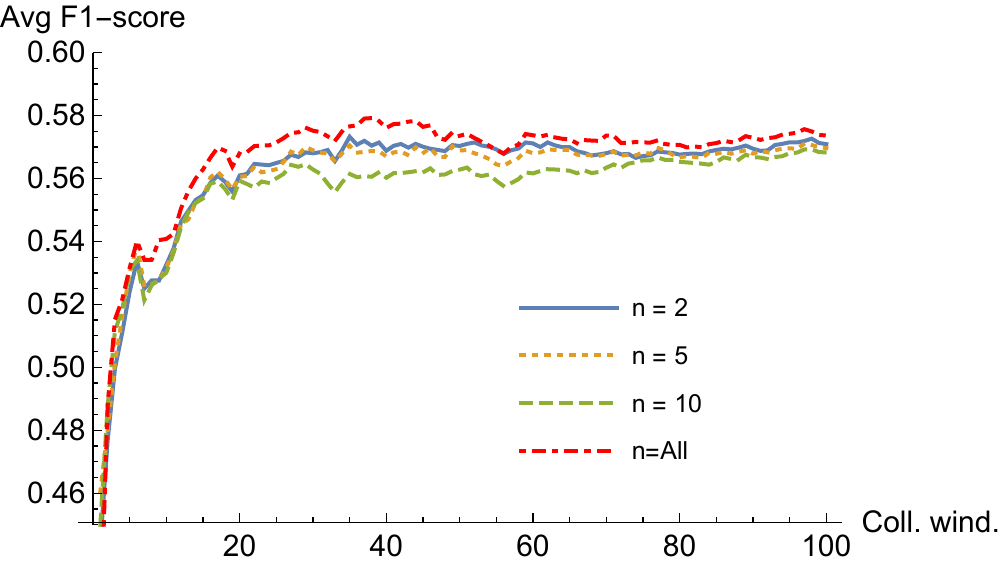}}
	\caption{Accuracy for different sizes of sample used by the re-training part. Uniform distribution of data on SMs.}
	\label{fig:complacc_unif}
\end{figure}

In Figures \ref{fig:complacc_zipf} and \ref{fig:complacc_unif} we report the accuracy results obtained for the Scenarios 2 and 3.
Each curve represents the accuracy of our approaches for different size of random samples used to train GreedyTL. As we can see, all the curves for both approaches are very close to each other. For the sake of comparison, we include in the plots also the accuracy curve obtained using all the local data to train GreedyTL (i.e., these curves are the same presented in the previous sections). Confidence intervals (95\% level of confidence)  have been computed but not shown they are too tight and would degrade the readability of the plots. 
In both scenarios we see that with SHTL the curves are very close to each other and most importantly, they are close to the one in which GreedyTL has been trained with all the local data. This means that we can obtain, as far as accuracy is concerned, the same results at a lower computational cost. Conversely,  A2AHTL appears to be less robust than SHTL. In fact, with $n=2$ in both scenarios the performance are inferior to the one corresponding to $n=5,10$. The reason is that, for some SMs, 2 data points per class are not enough to identify which source models are really informative.  

 To have a quantitative estimation of the performance degradation induced by the amount of data used, in  Tables \ref{tab:sc2complexacc} and \ref{tab:sc3complexacc} we report the mean accuracy values for all the approaches and sample sizes considered so far  and we compare them with the values reported in Sections~\ref{ssec:scenario2} and \ref{ssec:scenario3}. In the settings of Scenario 2 (i.e., where the data collection process in unbalances and few mule hold the vast majority of data) GreedyTL allows us to limit the amount of data used in the learning process losing up to 2\% of accuracy with respect to using all the data present in the local datasets. In Scenario 3, where data in uniformly distributed among mules, the loss of accuracy obtained by limiting the number of data observation used in the retraining phase is 3\%.
These results show empirically that, although at a relatively small cost, the proposed solutions can be, within certain limit, tuned to fit the computational requirements that might be imposed by a specific scenario. 

\begin{table}[ht]
	\centering
	\begin{tabular}{|c|c|c|c|c|}
		\hline

	 Algo.	& n=2 & n=5 & n=10 & n=All   \\
	\hline
	A2AHTL & 7\% & 6\% & 6\% & 5\% \\ 
	\hline
	StarHTL & 7\% & 7\% & 7\% & 6\% \\
	\hline
	\end{tabular}
\caption{Loss of accuracy compared to EdgeOnly for A2AHTL and StarHTL at different size of data used for retraining in Scenario 2 settings.}
\label{tab:sc2complexacc}
\end{table}

\begin{table}[ht]
	\centering
	\begin{tabular}{|c|c|c|c|c|}
		\hline
		
		 Algo.	& n=2 & n=5 & n=10 & n=All   \\
		\hline
		A2AHTL & 10\% & 8\% & 8\% & 7\% \\ 
		\hline
		StarHTL & 10\% & 8\% & 8\% & 8\% \\
		\hline
	\end{tabular}
	\caption{Loss of accuracy compared to EdgeOnly for A2AHTL and StarHTL at different size of data used for GreedyTL training in Scenario 3 settings. }
	\label{tab:sc3complexacc}
\end{table}

%% file: conclusions.tex
\section{Conclusions} \label{sec:conclusions}

In this paper we analysed the trade-off between accuracy and the energy consumed for collecting data from IoT sensors and performing a distributed learning task on such data.  We identified through simulations how to configure the learning process in order to reach a target loss of accuracy with respect to a centralised solution, minimising the amount of energy used to transfer data to the nodes involved in the learning process.  Our results show that, according to the network technologies we take into consideration it is possible to strongly limit the usage of cellular data transmission in favour of short-range wireless  technologies. In fact, being able to manage the data collection (and aggregation) between mobile devices, in such a way that they can collect enough data, proved to be the solution that allows us to save up to $94\%$ of energy resources (compared to sending all the data using a low energy cellular technology such as NB-IoT) and to obtain a model with accuracy comparable with a centralised cloud based solution, i.e. with only up to $2\%$ of loss of accuracy. Additionally, our approach is robust enough such that we can significantly control and reduce the computational impact of part our solution at a negligible cost in terms of accuracy (up to 3\% with respect to the plain version).

%% file: main.bbl
\begin{thebibliography}{10}
\expandafter\ifx\csname url\endcsname\relax
  \def\url#1{\texttt{#1}}\fi
\expandafter\ifx\csname urlprefix\endcsname\relax\def\urlprefix{URL }\fi
\expandafter\ifx\csname href\endcsname\relax
  \def\href#1#2{#2} \def\path#1{#1}\fi

\bibitem{Cisco:sp}
Cisco.
\newblock
  \href{http://www.cisco.com/c/en/us/solutions/collateral/service-provider/visual-networking-index-vni/mobile-white-paper-c11-520862.html}{Cisco
  visual networking index: Global mobile data traffic forecast update,
  2015--2020} [online] (2016).

\bibitem{McKinsey}
J.~Manyika, M.~Chui, J.~Bughin, R.~Dobbs, P.~Bisson, A.~Marrs.
\newblock \href{http://tinyurl.com/zpkr7m6}{Disruptive technologies: Advances
  that will transform life, business, and the global economy} [online]
  (McKinsey Global Institute, 2013).

\bibitem{Borgia20141}
E.~Borgia,
  \href{http://www.sciencedirect.com/science/article/pii/S0140366414003168}{The
  internet of things vision: Key features, applications and open issues},
  Computer Communications 54 (2014) 1 -- 31.
\newblock \href
  {https://doi.org/http://dx.doi.org/10.1016/j.comcom.2014.09.008}
  {\path{doi:http://dx.doi.org/10.1016/j.comcom.2014.09.008}}.
\newline\urlprefix\url{http://www.sciencedirect.com/science/article/pii/S0140366414003168}

\bibitem{ETSI-arch}
{ETSI TC M2M}.
\newblock \href{http://tinyurl.com/hemc3po}{{ETSI TS 102 690 v2.1.1 (2013-10)
  -- Machine-to-Machine Communications (M2M); Functional Architecture, 2013}}
  [online].

\bibitem{Kagermann:2013aa}
H.~Kagermann, W.~Wahlster, J.~Helbig, Recommendations for implementing the
  strategic initiative industrie 4.0, Tech. rep., German National Academy of
  Science and Engigeering (2013).

\bibitem{Bonomi:2012aa}
F.~Bonomi, R.~Milito, J.~Zhu, S.~Addepalli,
  \href{http://doi.acm.org/10.1145/2342509.2342513}{Fog computing and its role
  in the internet of things}, in: Proceedings of the First Edition of the MCC
  Workshop on Mobile Cloud Computing, MCC '12, ACM, New York, NY, USA, 2012,
  pp. 13--16.
\newblock \href {https://doi.org/10.1145/2342509.2342513}
  {\path{doi:10.1145/2342509.2342513}}.
\newline\urlprefix\url{http://doi.acm.org/10.1145/2342509.2342513}

\bibitem{Lopez:2015aa}
P.~G. Lopez, A.~Montresor, D.~Epema, A.~Datta, T.~Higashino, A.~Iamnitchi,
  M.~Barcellos, P.~Felber, E.~Riviere, Edge-centric computing: Vision and
  challenges, ACM Sigcomm Computer Communication Review (2015).

\bibitem{Satyanarayanan:2009aa}
M.~Satyanarayanan, P.~Bahl, R.~Caceres, N.~Davies, The case for vm-based
  cloudlets in mobile computing, IEEE Pervasive Computing 8~(4) (2009) 14--23.
\newblock \href {https://doi.org/10.1109/MPRV.2009.82}
  {\path{doi:10.1109/MPRV.2009.82}}.

\bibitem{Valerio:2016aa}
L.~Valerio, A.~Passarella, M.~Conti, Hypothesis transfer learning for efficient
  data computing in smart cities environments, in: International Conference on
  Smart computing (SMARTCOMP 2016), St. Louis, Missouri, 2016.

\bibitem{VALERIO201746}
L.~Valerio, A.~Passarella, M.~Conti,
  \href{http://www.sciencedirect.com/science/article/pii/S1574119217303875}{A
  communication efficient distributed learning framework for smart
  environments}, Pervasive and Mobile Computing 41 (2017) 46 -- 68.
\newblock \href {https://doi.org/https://doi.org/10.1016/j.pmcj.2017.07.014}
  {\path{doi:https://doi.org/10.1016/j.pmcj.2017.07.014}}.
\newline\urlprefix\url{http://www.sciencedirect.com/science/article/pii/S1574119217303875}

\bibitem{Valerio:2016ab}
L.~Valerio, A.~Passarella, M.~Conti, Accuracy vs. traffic trade-off of learning
  iot data patterns at the edge with hypothesis transfer learning, in: 2016
  IEEE 2nd International Forum on Research and Technologies for Society and
  Industry Leveraging a better tomorrow (RTSI), 2016, pp. 1--6.
\newblock \href {https://doi.org/10.1109/RTSI.2016.7740634}
  {\path{doi:10.1109/RTSI.2016.7740634}}.

\bibitem{Breiman:aa}
L.~Breiman, \href{http://dx.doi.org/10.1007/BF00058655}{Bagging predictors},
  Machine Learning 24~(2)  123--140.
\newblock \href {https://doi.org/10.1007/BF00058655}
  {\path{doi:10.1007/BF00058655}}.
\newline\urlprefix\url{http://dx.doi.org/10.1007/BF00058655}

\bibitem{Freund:1997aa}
Y.~Freund, R.~E. Schapire,
  \href{http://www.sciencedirect.com/science/article/pii/S002200009791504X}{A
  decision-theoretic generalization of on-line learning and an application to
  boosting}, Journal of Computer and System Sciences 55~(1) (1997) 119 -- 139.
\newblock \href {https://doi.org/http://dx.doi.org/10.1006/jcss.1997.1504}
  {\path{doi:http://dx.doi.org/10.1006/jcss.1997.1504}}.
\newline\urlprefix\url{http://www.sciencedirect.com/science/article/pii/S002200009791504X}

\bibitem{Freund:1997ab}
Y.~Freund, R.~E. Schapire, Y.~Singer, M.~K. Warmuth,
  \href{http://doi.acm.org/10.1145/258533.258616}{Using and combining
  predictors that specialize}, in: Proceedings of the Twenty-ninth Annual ACM
  Symposium on Theory of Computing, STOC '97, ACM, New York, NY, USA, 1997, pp.
  334--343.
\newblock \href {https://doi.org/10.1145/258533.258616}
  {\path{doi:10.1145/258533.258616}}.
\newline\urlprefix\url{http://doi.acm.org/10.1145/258533.258616}

\bibitem{Jacobs:1991aa}
R.~A. Jacobs, M.~I. Jordan, S.~J. Nowlan, G.~E. Hinton, Adaptive mixtures of
  local experts, Neural Computation 3~(1) (1991) 79--87.

\bibitem{Bordes:2005aa}
A.~Bordes, S.~Ertekin, J.~Weston, L.~Bottou,
  \href{http://dl.acm.org/citation.cfm?id=1046920.1194898}{Fast kernel
  classifiers with online and active learning}, J. Mach. Learn. Res. 6 (2005)
  1579--1619.
\newline\urlprefix\url{http://dl.acm.org/citation.cfm?id=1046920.1194898}

\bibitem{Dean:2012aa}
J.~Dean, G.~Corrado, R.~Monga, K.~Chen, M.~Devin, M.~Mao, A.~Senior, P.~Tucker,
  K.~Yang, Q.~V. Le, et~al., Large scale distributed deep networks, in:
  Advances in Neural Information Processing Systems, 2012, pp. 1223--1231.

\bibitem{Coates:2013aa}
A.~Coates, B.~Huval, T.~Wang, D.~Wu, B.~Catanzaro, A.~Ng, Deep learning with
  cots hpc systems, JLMR 28~(3) (2013) 1337--1345.

\bibitem{navia2006distributed}
A.~Navia-V{\'a}zquez, E.~Parrado-Hernandez, Distributed support vector
  machines, Neural Networks, IEEE Transactions on 17~(4) (2006) 1091--1097.

\bibitem{Georgopoulos20142}
L.~Georgopoulos, M.~Hasler,
  \href{http://www.sciencedirect.com/science/article/pii/S0925231213003639}{Distributed
  machine learning in networks by consensus}, Neurocomputing 124 (2014) 2 --
  12.
\newblock \href
  {https://doi.org/http://dx.doi.org/10.1016/j.neucom.2012.12.055}
  {\path{doi:http://dx.doi.org/10.1016/j.neucom.2012.12.055}}.
\newline\urlprefix\url{http://www.sciencedirect.com/science/article/pii/S0925231213003639}

\bibitem{Scardapane2015271}
S.~Scardapane, D.~Wang, M.~Panella, A.~Uncini,
  \href{http://www.sciencedirect.com/science/article/pii/S0020025515000298}{Distributed
  learning for random vector functional-link networks}, Information Sciences
  301 (2015) 271 -- 284.
\newblock \href {https://doi.org/http://dx.doi.org/10.1016/j.ins.2015.01.007}
  {\path{doi:http://dx.doi.org/10.1016/j.ins.2015.01.007}}.
\newline\urlprefix\url{http://www.sciencedirect.com/science/article/pii/S0020025515000298}

\bibitem{Konecny:2016aa}
J.~Kone{\v{c}}n{\`y}, H.~B. McMahan, F.~X. Yu, P.~Richt{\'a}rik, A.~T. Suresh,
  D.~Bacon, Federated learning: Strategies for improving communication
  efficiency, arXiv preprint arXiv:1610.05492 (2016).

\bibitem{Oquab:2014aa}
M.~Oquab, L.~Bottou, I.~Laptev, J.~Sivic, Learning and transferring mid-level
  image representations using convolutional neural networks, in: Computer
  Vision and Pattern Recognition (CVPR), 2014 IEEE Conference on, IEEE, 2014,
  pp. 1717--1724.

\bibitem{Yosinski:2014aa}
J.~Yosinski, J.~Clune, Y.~Bengio, H.~Lipson, How transferable are features in
  deep neural networks?, in: Advances in neural information processing systems,
  2014, pp. 3320--3328.

\bibitem{Valerio:2017aa}
L.~Valerio, A.~Passarella, M.~Conti,
  \href{https://doi.org/10.1109/WoWMoM.2017.7974310}{Optimal trade-off between
  accuracy and network cost of distributed learning in mobile edge computing:
  An analytical approach}, in: 18th {IEEE} International Symposium on {A} World
  of Wireless, Mobile and Multimedia Networks, WoWMoM 2017, Macau, China, June
  12-15, 2017, 2017, pp. 1--9.
\newblock \href {https://doi.org/10.1109/WoWMoM.2017.7974310}
  {\path{doi:10.1109/WoWMoM.2017.7974310}}.
\newline\urlprefix\url{https://doi.org/10.1109/WoWMoM.2017.7974310}

\bibitem{nbiot:2016aa}
\href{http://www.3gpp.org/news-events/3gpp-news/1785-nb\_iot\_complete}{Nb-iot}
  (2016).
\newline\urlprefix\url{http://www.3gpp.org/news-events/3gpp-news/1785-nb\_iot\_complete}

\bibitem{zigbee:aa}
\href{https://www.zigbee.org}{Zigbee} (2018).
\newline\urlprefix\url{https://www.zigbee.org}

\bibitem{OrabonaGreedyTL}
I.~Kuzborskij, F.~Orabona, B.~Caputo, Transfer learning through greedy subset
  selection, in: Proceedings of Intl. Conf. on Image Analysis and Processing,
  2015.

\bibitem{Jensen:2012aa}
A.~R. Jensen, M.~Lauridsen, P.~Mogensen, T.~B. S{\o}rensen, P.~Jensen, Lte ue
  power consumption model: For system level energy and performance
  optimization, in: 2012 IEEE Vehicular Technology Conference (VTC Fall), 2012,
  pp. 1--5.
\newblock \href {https://doi.org/10.1109/VTCFall.2012.6399281}
  {\path{doi:10.1109/VTCFall.2012.6399281}}.

\bibitem{Sinha:2017aa}
R.~S. Sinha, Y.~Wei, S.-H. Hwang, A survey on lpwa technology: Lora and nb-iot,
  ICT Express 3~(1) (2017) 14--21.

\bibitem{Ahmed:2016aa}
N.~Ahmed, H.~Rahman, M.~Hussain,
  \href{http://www.sciencedirect.com/science/article/pii/S2405959516300650}{A
  comparison of 802.11ah and 802.15.4 for iot}, ICT Express 2~(3) (2016) 100 --
  102, special Issue on ICT Convergence in the Internet of Things (IoT).
\newblock \href {https://doi.org/https://doi.org/10.1016/j.icte.2016.07.003}
  {\path{doi:https://doi.org/10.1016/j.icte.2016.07.003}}.
\newline\urlprefix\url{http://www.sciencedirect.com/science/article/pii/S2405959516300650}

\bibitem{Gomez:2013aa}
K.~Gomez, T.~Rasheed, R.~Riggio, D.~Miorandi, C.~Sengul, N.~Bayer, Achilles and
  the tortoise: Power consumption in ieee 802.11n and ieee 802.11g networks,
  in: 2013 IEEE Online Conference on Green Communications (OnlineGreenComm),
  2013, pp. 20--26.
\newblock \href {https://doi.org/10.1109/OnlineGreenCom.2013.6731023}
  {\path{doi:10.1109/OnlineGreenCom.2013.6731023}}.

\bibitem{Powers:2011aa}
D.~M.~W. Powers, {Evaluation: From precision, recall and f-measure to roc.,
  informedness, markedness \& correlation}, Journal of Machine Learning
  Technologies 2~(1) (2011) 37--63.

\bibitem{Arnaboldi:2017aa}
V.~Arnaboldi, M.~G.~G. Campana, F.~Delmastro, Context-aware configuration and
  management of wifi direct groups for real opportunistic networks, in: 2017
  IEEE 14th International Conference on Mobile Ad Hoc and Sensor Systems
  (MASS), 2017, pp. 266--274.
\newblock \href {https://doi.org/10.1109/MASS.2017.40}
  {\path{doi:10.1109/MASS.2017.40}}.

\bibitem{Conti:2013aa}
M.~Conti, F.~Delmastro, G.~Minutiello, R.~Paris, Experimenting opportunistic
  networks with wifi direct, in: 2013 IFIP Wireless Days (WD), 2013, pp. 1--6.
\newblock \href {https://doi.org/10.1109/WD.2013.6686501}
  {\path{doi:10.1109/WD.2013.6686501}}.

\bibitem{Shalev-Shwartz:2008aa}
S.~Shalev-Shwartz, N.~Srebro,
  \href{http://doi.acm.org/10.1145/1390156.1390273}{Svm optimization: Inverse
  dependence on training set size}, in: Proceedings of the 25th International
  Conference on Machine Learning, ICML '08, ACM, New York, NY, USA, 2008, pp.
  928--935.
\newblock \href {https://doi.org/10.1145/1390156.1390273}
  {\path{doi:10.1145/1390156.1390273}}.
\newline\urlprefix\url{http://doi.acm.org/10.1145/1390156.1390273}

\bibitem{Kuzborskij:2017aa}
I.~Kuzborskij, F.~Orabona, B.~Caputo,
  \href{http://www.sciencedirect.com/science/article/pii/S1077314216301370}{Scalable
  greedy algorithms for transfer learning}, Computer Vision and Image
  Understanding 156 (2017) 174 -- 185, image and Video Understanding in Big
  Data.
\newblock \href {https://doi.org/https://doi.org/10.1016/j.cviu.2016.09.003}
  {\path{doi:https://doi.org/10.1016/j.cviu.2016.09.003}}.
\newline\urlprefix\url{http://www.sciencedirect.com/science/article/pii/S1077314216301370}

\end{thebibliography}
